\documentclass[lettersize,journal]{IEEEtran}
\usepackage{amsmath,amsfonts}
\usepackage{amssymb}
\usepackage{algorithmic}
\usepackage{algorithm}
\usepackage{array}
\usepackage{textcomp}
\usepackage{stfloats}
\usepackage{url}
\usepackage{verbatim}
\usepackage{xcolor}
\usepackage{graphicx}
\usepackage{balance}
\usepackage{cite}
\usepackage{subcaption}
\usepackage{float}
\usepackage[export]{adjustbox}
\usepackage{wrapfig}
\usepackage{multirow}
\usepackage{hyperref}
\usepackage{mathtools,lipsum}

\hypersetup{hidelinks}

\newtheorem{remark}{Remark}

\newcommand{\HH}{\mathbf{H}}

\newcommand{\FF}{\mathbf{F}}

\newcommand{\hermconj}{^{\mathsf{H}}}
\newcommand\norm[1]{\left\lVert#1\right\rVert}

\begin{document}

\title{Impact of Synchronization Offsets and CSI Feedback Delays on Coherent Joint Transmission in Distributed MIMO-OFDM}

\author{Kumar Sai Bondada,~\IEEEmembership{Student Member,~IEEE,}
        Daniel J. Jakubisin,~\IEEEmembership{Senior Member,~IEEE,} 
        and \\ R. Michael Buehrer,~\IEEEmembership{Fellow,~IEEE}
        
\thanks{Kumar Sai Bondada, R. Michael Buehrer, and Daniel J. Jakubisin are with Wireless@Virginia Tech and the Bradley Department of Electrical and Computer Engineering, Virginia Tech, Blacksburg, VA, USA. 
Daniel J. Jakubisin is also with the Hume Center for National Security and Technology, Virginia Tech. 
Corresponding author: Kumar Sai Bondada (kumarsaibondada@vt.edu).}

\vspace{-8mm}
}




\maketitle


\begin{abstract}
This paper studies the impact of synchronization offsets and CSI feedback delay on coherent joint transmission (CJT) in distributed MIMO-OFDM systems. Unlike conventional co-located MIMO, distributed CJT is highly sensitive to transmitter-dependent synchronization impairments because coherent combining requires phase, frequency, and timing alignment across geographically separated transmitters. Received signal models are derived for residual timing offsets, carrier frequency offsets (CFOs), sampling clock offsets (SCOs), phase noise, and channel variations accumulated during channel state information (CSI) feedback delay. The corresponding spectral efficiency (SE) expressions quantify the degradation in capacity due to loss in coherent-combining gain caused by residual phase rotations, inter-carrier interference (ICI), and precoder misalignment. The SE expressions are numerically evaluated to determine synchronization requirements for maintaining negligible SE degradation, yielding the following findings: a Gaussian-distributed start-time jitter with a standard deviation below 10\% of the sampling period results in negligible SE degradation, with the SE loss decreasing from 0.68\% at an SNR of 0 dB to 0.17\% at 40 dB. A normalized CFO standard deviation of 0.001 also results in negligible ICI-induced degradation, with the SE loss increasing from 0.05\% at an SNR of 0 dB to 2\% at 40 dB. At an SNR of 26 dB, limiting the SE loss to 5\% yields feasible operating points with clock-error standard deviations up to 50 ppb, Doppler-delay products up to 0.1, and phase-tracking error standard deviations up to $40^{\circ}$ at carrier frequency of 3.5 GHz with 30 kHz sub-carrier spacing with 15.36 MHz sampling rate.
\end{abstract}


\vspace{-1.5em}
\section{Introduction}
\IEEEPARstart{D}{istributed} multiple-input multiple-output (DMIMO) architectures, which employ a large number of antennas distributed over a geographical area, have been extensively studied~\cite{6495775, 6574665,9113273,9492307,Said_MDMMIMO_IEEEWCM} and have shown significant performance gains compared with conventional co-located MIMO. These gains stem from shorter access distances and relaxed form-factor constraints allowing the virtual antenna array size to scale, thereby improving the quality of service (QoS), spectral efficiency (SE), and energy efficiency (EE). Prior work has proposed various DMIMO architectures under different objectives, including the analysis of the SE of single-user (SU) and multi-user (MU) MIMO \cite{5490981,5953530}, inter-cell interference mitigation \cite{5706317}, uniform QoS provisioning across the network \cite{7827017}, and extending communication range \cite{9861593}. Existing studies have also investigated enhancements to DMIMO architectures, including the connection of distributed antennas to a central processing unit through wired links, use of wireless fronthaul links, and mobility support~\cite{7938406,9615200,10757642,11036858}.

Coordinated multi-point (CoMP) transmission in LTE~\cite{Marsch_Fettweis_2011} and the multiple transmit/receive point (multi-TRP) framework in 5G~\cite{3GPP_TS_38_300_V19_4_0,10121037} represent standardized cellular implementations of DMIMO designed to improve interference management, coverage, reliability, and cell-edge throughput. In Release~19 of 3GPP TS~38.300, 3GPP specifies UE-assisted inter-TRP calibration for coherent joint transmission (CJT)~\cite{3GPP_TS_38_300_V19_4_0}. Specifically, each TRP transmits a channel state information reference signal (CSI-RS), enabling the UE to estimate the relative delay, frequency, and phase offsets among the received TRP signals and report these quantities to the network through a CSI report carried on the physical uplink shared channel (PUSCH). Given the stringent data-rate targets envisioned for 6G~\cite{ITU2024}, CJT is a promising technique because of its potential to provide substantial SE gains. Cell-free (CF) massive MIMO~\cite{11368664,7827017,10694465} follows a related distributed-cooperation architecture, in which distributed transceivers jointly serve UEs through coordinated precoding and combining without conventional cell boundaries.

Although DMIMO systems offer significant performance improvements, they face several practical challenges, including stringent synchronization requirements, accurate CSI acquisition, and fronthaul capacity and latency constraints~\cite{9492307,6574665}. Prior work has explored several synchronization strategies, including wired, GPS-based, and GPS-independent wireless approaches. Recent wired solutions include the White Rabbit (WR) protocol, an extension of the Precision Time Protocol (PTP) that incorporates Synchronous Ethernet (SyncE) for syntonization and phase detectors for phase correction. WR achieves sub-ns timing accuracy and high relative frequency coherence, with ps-level integrated jitter~\cite{6070148}. However, wired solutions limit mobility, scalability, and adaptability.

In GPS-based approaches, a GPS-disciplined oscillator (GPSDO), comprising a GPS receiver and a high-quality local oscillator (LO), provides stable timing and frequency references. GPSDOs can achieve timing accuracies on the order of tens of ns and fractional frequency stability on the order of $0.1$--$10$~parts per trillion (ppt)~\cite{article_Lombardi}, with excellent long-term stability. Their short-term stability is mainly determined by the LO phase noise characteristics. Although GPSDOs enable synchronization without wired connectivity, they are vulnerable to GPS spoofing and intentional interference, increase power consumption and cost, and may degrade under non-line-of-sight (NLOS) conditions. Also, GPS coverage is not ubiquitous.
To eliminate GPS dependency, GPS-independent wireless approaches have been developed and can be classified into centralized~\cite{Liu2026DoubleLayer, Li2024CooperativeTime,Merlo2026RealTimeCalibration,7218555,bondada2025experimentaldemonstrationrobustdistributed} and decentralized~\cite{9173801,10734842}. In centralized approaches, followers synchronize to a leader through wireless frequency and time reference exchanges. Frequency synchronization is typically achieved by transmitting an analog~\cite{7218555,Merlo2026RealTimeCalibration,bondada2025experimentaldemonstrationrobustdistributed } or digital preamble-based reference~\cite{Liu2026DoubleLayer}, whereas time synchronization commonly relies on a two-way time-transfer technique~\cite{Merlo2026RealTimeCalibration,Li2024CooperativeTime}. These methods have demonstrated sub-Hz frequency accuracy and ns-ps timing accuracy. However, their performance is sensitive to multipath, Doppler, propagation-delay asymmetry, and the SNR of the wireless synchronization link. 
\vspace{-1.0em}
\subsection{Related Work and Motivation}
Despite these advances, synchronization offsets remain unavoidable in practical deployments, and their impact on CJT performance in DMIMO systems after synchronization correction remains insufficiently understood. A limited number of studies~\cite{1417557,4202181,4533899,5191056} have investigated synchronization errors in DMIMO systems, often considering timing or frequency errors in isolation and evaluating their impact through bit/symbol-error-rate (BER/SER) analysis. The study in~\cite{1417557} evaluated the impact of clock jitter, which causes inter-symbol interference (ISI), on transmit maximal-ratio combining (MRC) and Alamouti space-time block code (STBC) schemes in a single-carrier setting. The results showed that a timing jitter of $10\%$ is tolerable without noticeable BER degradation. The work in~\cite{4202181} examined the effects of initial phase-offset errors, assuming ideal time alignment, in a single-antenna multiple-transmitter scenario. It concluded that beamforming gains remain high for phase errors up to $60^{\circ}$ and proposed a leader-follower synchronization protocol, in which the leader broadcasts a reference carrier and the followers correct their phases by measuring round-trip phase offsets.

The work in~\cite{4533899} demonstrated that timing offsets cause ISI, destroy the orthogonality of STBC codes, and degrade channel estimation (CE) accuracy, with synchronization errors of up to $20\%$ of the sampling period being tolerable. The study in~\cite{5191056} showed that, as the number of cooperating transceivers increases, ISI due to timing offsets also increases, thereby degrading the cooperative gains. These studies provide valuable insights either in improving performance of synchronization techniques or providing bounds. \textit{However, further study is needed to characterize the effects of synchronization errors and CSI aging on the SE of DMIMO systems, particularly in multicarrier settings. This is important for quantifying synchronization requirements and enabling synchronization-aware precoder design in practical CJT systems.} Accordingly, our study supports ongoing development and standardization efforts for CJT, including 3GPP's work on UE-assisted inter-TRP calibration~\cite{3GPP_TS_38_300_V19_4_0} and research toward multi-AP (Access Point) joint transmission in future Wi-Fi systems~\cite{Levinbook2024}. Additionally, our findings provide guidance on the accuracy requirements for distributed synchronization techniques to enable effective CJT.

\vspace{-1.0em}
\subsection{Contribution}



Our prior work~\cite{11162345} investigated the SE performance of CJT under residual timing and frequency offsets and CSI feedback delay in a narrowband, single-carrier Rayleigh fading setting. Building on that work, this paper develops a wideband distributed SU MIMO-OFDM framework over multipath Rayleigh fading channels. The framework captures multicarrier-specific effects, including the valid timing-window condition for processing received signals from transmitters at the receiver, timing jitter, CFO- and SCO-induced ICI and phase evolution, SCO-induced timing drift, and phase-estimation errors associated with synchronization-aware precoder compensation. The analysis proceeds progressively, beginning with an ideal-synchronization benchmark and sequentially incorporating transmitter-dependent impairments to characterize their effects on CJT.

The main contributions are as follows: 1) we derive the received signal model and SE expression for distributed SU-MIMO-OFDM CJT under ideal synchronization, which serve as a baseline for quantifying synchronization-induced degradation; 2) we extend the model to residual timing offsets and CFOs, identify the valid receiver processing window within the cyclic prefix (CP), and characterize CFO-induced phase rotations, desired-signal attenuation, and ICI; 3) we incorporate SCOs and show that they cause subcarrier- and symbol-dependent phase evolution and timing drift across OFDM symbols, thereby motivating periodic timing realignment; 4) \textit{we derive a phase-compensation term for the precoder} and quantify the degradation caused by imperfect phase compensation; 5) we analyze phase noise and CSI feedback delay, showing that phase fluctuations and CSI aging cause precoder-channel misalignment and reduce the coherent-combining gain; 6) we numerically evaluate the derived SE expressions through Monte Carlo averaging over independent channel and synchronization-offset realizations to quantify the synchronization requirements; and \textcolor{black}{finally, 7) we provide ``rules of thumb" about the required accuracy relative to the parameters such as CP duration, subcarrier spacing, sampling period.}

 \textit{Assumptions:} We assume that the transmit data and precoder information are available at all cooperating transmitters. To isolate the impact of synchronization impairments on CJT, large-scale path loss and shadowing are not explicitly modeled, and the links from the cooperating transmitters are normalized to have the same average received SNR. Path loss and network geometry primarily determine the operating SNR and the relative strengths of the distributed links. In contrast, the objective of this work is to determine whether the coherent-combining gain can be preserved once a given received-SNR operating point is achieved. Incorporating path loss, transmitter selection, and power control is left for future work, where network geometry becomes central to the analysis. 

\textit{Outline:} The paper is organized as follows. Section~\ref{sec:system_model} presents the system model and derives the SE expression for distributed SU MIMO-OFDM CJT under ideal synchronization. Section~\ref{ref:se_section_offsets} extends the analysis to synchronization offsets and CSI feedback delay. Section~\ref{sec:simulation_results_and_analysis} presents numerical evaluation of the SE expressions to quantify the impact of these impairments on SE and determine synchronization requirements. Finally, Section~\ref{sec:conclusion_and_future_work} concludes the paper and discusses future work.


        

\section{Distributed Single-User MIMO-OFDM} \label{sec:system_model}
\noindent We consider a DMIMO-OFDM system with $N_T$ geographically separated transmitters jointly serving a receiver. Each transmitter is equipped with $N_t$ antennas, while the receiver is equipped with $N_r$ antennas. Each OFDM symbol contains $K$ subcarriers and a cyclic prefix (CP) of duration $T_{\mathrm{cp}}$, corresponding to $K_{\mathrm{cp}}$ samples. Thus, the total number of samples per OFDM symbol is $K_0=K+K_{\mathrm{cp}}$. The useful OFDM symbol duration is denoted by $T_{\mathrm{eff}}=1/f_{\mathrm{scs}}$, where $f_{\mathrm{scs}}$ is the subcarrier spacing, and the total OFDM symbol duration is $T_0=T_{\mathrm{eff}}+T_{\mathrm{cp}}$. The signal transmitted from the $m$th antenna $(m\in\{1,\ldots,N_t\})$ of the $q$th transmitter $(q\in\{1,\ldots,N_T\})$ over a block of $L$ consecutive OFDM symbols is given by 
\begin{equation*} 
x_q^m(t) = \sum_{l=0}^{L-1} s_q^{m,l}(t-lT_0), 
\vspace{-.25em}
\end{equation*}
where 
\begin{equation*} s_q^{m,l}(t) = \sum_{k=-K/2}^{K/2-1} s_{q,k}^{m,l} e^{j2\pi k(t-T_{\mathrm{cp}})/T_{\mathrm{eff}}}, \quad 0\le t< T_0,
\vspace{-.3em}
\end{equation*}
$s_q^{m,l}(t) = s_q^{m,l}(t+T_{\mathrm{eff}}),\,0\le t < T_{\mathrm{cp}}$ and $s_{q,k}^{m,l}$ denotes the precoded data or pilot symbol transmitted on the $k$th subcarrier during the $l$th OFDM symbol. The received signal at the $n$th antenna of the receiver, where
$n\in\{1,\ldots,N_r\}$, is the superposition of the signals transmitted
from all distributed transmit antennas after propagation through their
respective channels. It is given by
\begin{align}
    & y^{n}(t)
    =
    \sum_{q=1}^{N_T}
    \sum_{m=1}^{N_t}
    \int_{0}^{\tau_{q}^{{\max}}}
    h_{q}^{n,m}(\tau)
    x_{q}^{m}(t-\tau)
    d\tau
    +w^{n}(t),
    \label{eq:received_signal_time}
\end{align}
where $h_{q}^{n,m}(\tau)$ denotes the time-invariant channel impulse
response between the $m$th antenna of transmitter $q$ and the $n$th
receive antenna, and $w^{n}(t)$ denotes additive white complex Gaussian noise (AWGN).
In the ideal case, it is assumed that the maximum channel delay spread $\tau_{q}^{{\max}}$ is shorter than the CP duration and that the channel remains time-invariant between the pilot transmission used for precoder design, occurring at the $(l-\mathcal{L})$th OFDM symbol, and the CJT at the $l$th OFDM symbol for initial synchronization error analysis. Thus,
$
    h_{q}^{n,m,l-\mathcal{L}}(\tau)
    =
    h_{q}^{n,m,l}(\tau).
$
Later, this assumption is relaxed to study the Doppler-induced CSI aging impact. For each received OFDM symbol, the receiver removes the CP and applies a bank of $K$ matched filters over the useful symbol interval, with one filter matched to each subcarrier, to extract the corresponding frequency-domain data symbols, i.e. Fourier transform, as derived in Appendix~\ref{Derivation1}. 

From~\eqref{eq:rx_signal_subcarrier}, the received signal across all
$N_r$ antennas of receiver on subcarrier $k$ during the $l$th OFDM
symbol can be expressed in matrix form as
\begin{align}
\vspace{-.25em}
    \mathbf{z}_{k}^{l}
    &=
    \mathbf{H}_{k}^{l}\mathbf{s}_{k}^{l}
    +
    \mathbf{w}_{k}^{l},
    \label{eq:rx_signal_subcarrier_matrix}
\end{align}
where $\mathbf{z}_{k}^{l}\in\mathbb{C}^{N_r\times 1}$ is the received
signal vector, $\mathbf{H}_{k}^{l}\in
\mathbb{C}^{N_r\times N_TN_t}$ is the composite channel matrix,
$\mathbf{s}_{k}^{l}\in\mathbb{C}^{N_TN_t\times 1}$ is the jointly transmitted
data signal vector, and $\mathbf{w}_{k}^{l}\in\mathbb{C}^{N_r\times 1}$ is
the additive-noise vector.
The jointly transmitted signal vector is precoded as
\begin{equation*}
    \mathbf{s}_k^l
    =
    \mathbf{F}_k^l \mathbf{d}_k^l,
    \qquad
    \mathbf{F}_k^l
    =
    \sqrt{\frac{P_k^l}{N_t}}\,\mathbf{W}_k^l,
\vspace{-.2em}
\end{equation*}
where
$
    \mathbf{W}_k^l
    =
    \begin{bmatrix}
        {\mathbf{W}_{1,k}^{l}}^\top &
        \cdots &
        {\mathbf{W}_{q,k}^{l}}^\top &
        \cdots &
        {\mathbf{W}_{N_T,k}^{l}}^\top
    \end{bmatrix}^{\top}
$
is the composite precoder, and
$\mathbf{W}_{q,k}^l \in \mathbb{C}^{N_t \times N_s}$
is the individual precoder applied at transmitter $q$.
Furthermore,
$\mathbf{d}_k^l \in \mathbb{C}^{N_s \times 1}$
is the common data-symbol vector transmitted by all cooperating
transmitters, and $P_k^l$ denotes the transmit power allocated
to each transmitter on subcarrier $k$ during OFDM symbol $l$.
The data symbol vectors are modeled as independent zero-mean, unit-covariance proper complex Gaussian random vectors, i.e.,
$
    \mathbf{d}_k^l \sim \mathcal{CN}\!\left(\mathbf{0},\mathbf{I}\right).
$
Accordingly,
\vspace{-.5em}
\begin{equation*}
    \mathbb{E}\!\left[
        \mathbf{d}_i^l
        {\mathbf{d}_k^l}^{\hermconj}
    \right]
    =
    \begin{cases}
        \mathbf{I}, & i=k,\\
        \mathbf{0}, & i\neq k.
    \end{cases}
\end{equation*}
This Gaussian signaling assumption is capacity-achieving for Gaussian channels under the specified covariance constraint. The entries of $\mathbf{H}_k$ are modeled as independent and identically distributed zero-mean, unit-variance proper complex Gaussian random variables across the transmitter-receiver antenna links, such that $\mathbb{E}\!\left[\lVert\mathbf{H}_k\rVert_F^2\right]=N_TN_tN_r$. This signal model assumes ideal timing and frequency synchronization, identical sampling clocks across the transmitters, and no oscillator phase noise. In addition, each transmitter is assumed to compensate for its corresponding propagation delay by advancing its transmission so that the signals from all transmitters arrive simultaneously at the receiver. 

Under ideal synchronization, the average achievable SE over a block of
$L$ OFDM symbols is
\begin{equation}
    {R}_{\mathrm{ideal}}= \frac{\kappa}{LK}
    \sum_{l=0}^{L-1}
    \sum_{k=-\frac{K}{2}}^{\frac{K}{2}-1}
    \log_2
    \det\left(
        \mathbf{I}
        +
        \mathbf{\Lambda}^l_{k}
    \right)
    \,\, \text{(bit/s/Hz)},
    \label{eq:ideal_se}
        \vspace{-.5em}
\end{equation}
where
$
    \mathbf{\Lambda}_{k}^l
    =
    \frac{1}{\sigma^2}
    \mathbf{H}_{k}^{l}\FF_{k}^{l}
    \left(
        \mathbf{H}_{k}^{l}\FF_{k}^{l}
    \right)^{\hermconj}.
$
The term $\kappa=\frac{T_{\mathrm{eff}}}{T_{0}}$ represents the overhead due to CP. For each transmitter $q$, the total transmit power allocated across all
subcarriers during OFDM symbol $l$ satisfies
\begin{equation}
    \sum_{k=-K/2}^{K/2-1} P_k^l \leq P,
    \label{eq:power_per_subcarrier}
    \vspace{-.5em}
\end{equation}
where $P$ denotes the per-transmitter power, assumed to be the same
for all transmitters.  
For simplicity, equal power allocation across subcarriers is considered, such that
$
P_k^l={P}/{K}.
$


SVD-based precoding is considered. For subcarrier $k$ and OFDM
symbol $l$, the composite channel is decomposed as
$
    \HH_k^l
    =
    \mathbf{U}_k^l
    \mathbf{\Sigma}_k^l
    {\mathbf{V}_k^l}^{\hermconj}.
$
The precoder is constructed from the first
$N_s=\min(N_TN_t,N_r)$ right singular vectors as
$
    \mathbf{W}_k^l
    =
    \mathbf{V}_{k,N_s}^l
    \sqrt{\boldsymbol{\Pi}_k^l},,
$
where $\mathbf{V}_{k,N_s}^l$ contains the beamforming directions
associated with the $N_s$ strongest singular modes, and
$
    \boldsymbol{\Pi}_k^l,
    =
    \operatorname{diag}
    \left(
        p_{k,1}^l,\ldots,p_{k,N_s}^l
    \right)
$
is the stream-power allocation matrix. Here, $p_{k,i}^l$ denotes
the normalized power allocated to the $i$th spatial stream. In the
full-power case, the stream-power allocation satisfies
\begin{equation*}
    \sum_{i=1}^{N_s}p_{k,i}^l=N_TN_t, \,
    \left(p_{k,i}^l=\frac{N_TN_t}{N_s}, \text{equal power allocation}\right).
\end{equation*} This normalization ensures that the average transmit power per
transmitter on subcarrier $k$ is $P_k^l$. In particular, we
can see that
\begin{equation}
\sum_k\frac{P_k^l}{N_t}
\mathbb{E}_{\mathbf{H}}\!\left[
\operatorname{tr}\!\left(
\mathbf{W}_{q,k}^{l}
{\mathbf{W}_{q,k}^{l}}^{\hermconj}
\right)
\right]
\approx P,
\qquad \forall q,
\end{equation}
where the expectation is taken over the channel realizations.
The SNR improvement arises from both the alignment of the precoder with the channel and the additional transmit power contributed by
the cooperating transmitters. To isolate the coherent array gain,
the total transmit power is normalized so that it remains constant
as the number of transmitters increases. Accordingly, the
stream-power allocation satisfies
$
    \sum_{i=1}^{N_s}p_{k,i}^l=N_t.
$

\vspace{-1.0em}
\section{Spectral Efficiency in presence of Impairments} 
\label{ref:se_section_offsets}
\vspace{-.5em}
\noindent Oscillators used as internal frequency references ideally generate perfectly stable sinusoidal signals. In practice, hardware non-idealities introduce phase noise and small frequency inaccuracies. Thus, the oscillator output can be modeled as
\vspace{-.25em}
\begin{equation}
    A(t)\cos\left(2\pi (f_o+\delta f)t+\nu(t)+\theta\right),
    \label{eq:oscillator_output}
\end{equation}
where $A(t)$ denotes the slowly varying amplitude (assumed to be constant), $f_o$ is the nominal reference frequency, $\delta f$ is the frequency offset, $\nu(t)$ models the phase noise process, and $\theta$ is the initial phase. The frequency offset arises from initial oscillator tolerance, temperature variations, aging, and supply or load fluctuations, resulting in a long-term frequency bias. In contrast, the phase noise process $\nu(t)$ captures short-term oscillator instability and is commonly modeled as a Wiener process with zero mean, $\nu(0)=0$, and independent Gaussian increments whose variance increases with the time interval~\cite{381848}.


The reference signal in~\eqref{eq:oscillator_output} drives both the carrier phase-locked loop (PLL), which generates the up-conversion and down-conversion carriers, and the clock PLL, which generates the ADC/DAC sampling clocks. As a result, the frequency offset in the reference oscillator appears as both carrier frequency offset (CFO) and sampling clock offset (SCO). In addition, PLL components, including the loop filter, voltage-controlled oscillator (VCO), frequency divider, and phase-frequency detector (PFD), add phase noise, which combines with reference-oscillator noise to determine the carrier phase noise and sampling clock jitter at the transceiver. Oscillator data sheets typically specify long-term frequency accuracy in parts per billion (ppb) or parts per million (ppm), along with phase noise levels in dBc/Hz at different offset frequencies. Each transceiver also maintains a local time reference, which may be aligned to GPS, UTC, or another local reference. Residual differences among the transmitter clock references, together with errors in estimating and compensating for propagation delays when advancing the transmission times (i.e., knowing when to start transmissions), result in residual time offsets at the receiver. 

In the following analysis, the residual offsets are studied separately to isolate their impact on coherent joint transmission (CJT). The residual offsets considered here are those remaining after synchronization correction. Their levels depend on the quality of the synchronization link and on the accuracy of the associated estimation and correction algorithms. The received signal in the presence of residual time and frequency offsets can be written as
\begin{align}
    y^{n}(t)
    &=
    \sum_{q=1}^{N_T} \sum_{m=1}^{N_t}
    e^{j2\pi \delta_q t }
    \left(x_q^m(t) * h_{q}^{n,m}(\tau)\right)
    + w^n(t),
    \label{eq:received_signal_offsets}
\end{align}
\noindent where
$    x_q^m(t) * h_{q}^{n,m}(\tau)
    =
    e^{j\varphi_q}\int_{0}^{\tau^{\max}_{q}}
    h_{q}^{n,m}(\tau)
    x_q^m(t-\tau-\Delta_q)
    d\tau .
$
Here, $\Delta_q$ is the residual time offset and $\delta_q$ is the residual frequency offset at the $q$th transmitter, respectively. The constant phase factor per transmitter $e^{j\varphi_q}$ is absorbed into the channel $h_{q}^{n,m}(\tau)$, which includes initial carrier phase offset for convenience. Appendix \ref{Derivation2} shows the derivation of the received signal model in the presence of residual time and frequency offsets.

\vspace{-1.00em}
\subsection{Time Synchronization Offsets}
\begin{figure}
    \centering
    \includegraphics[width=0.95\linewidth]{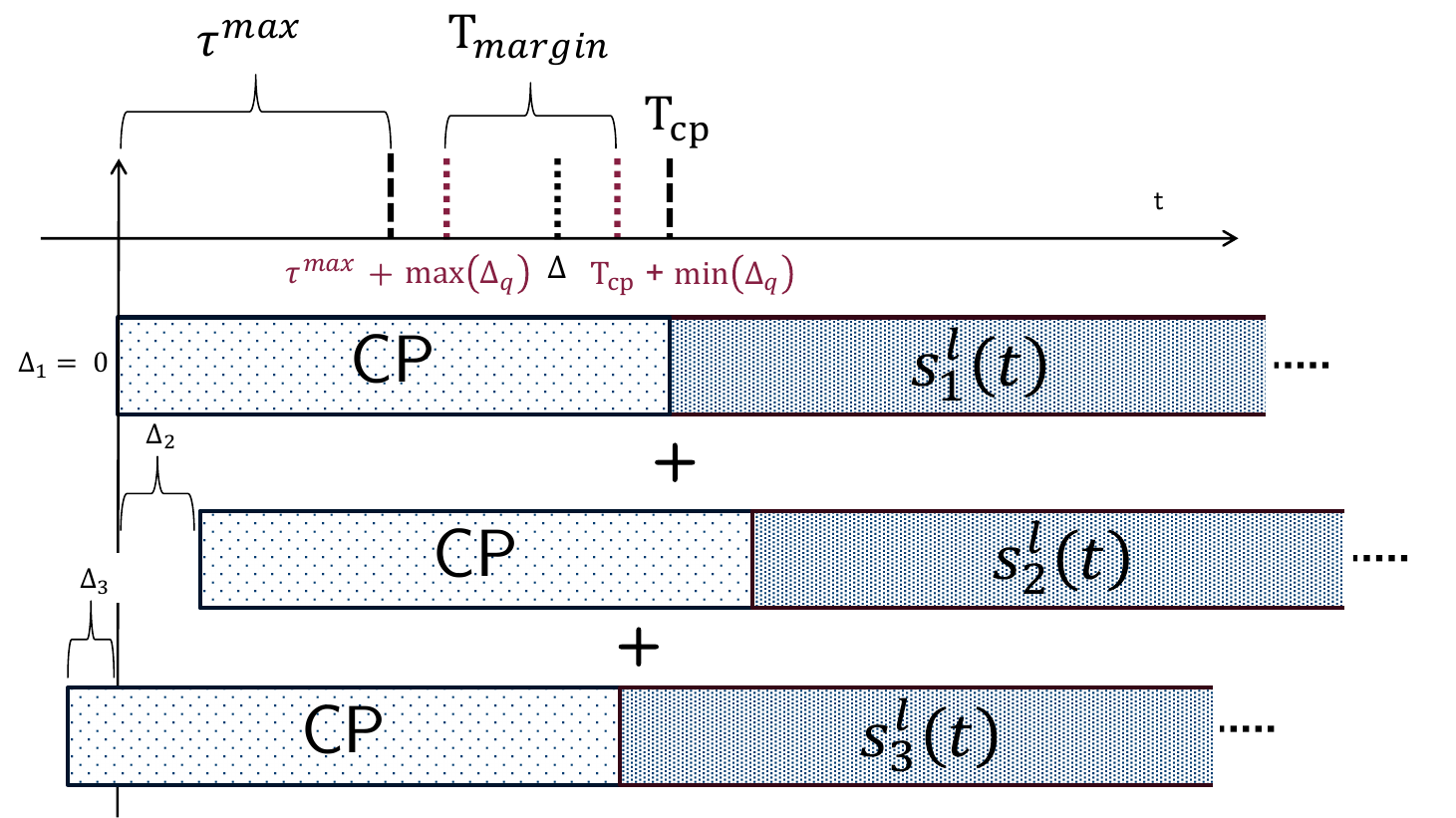}
    \caption{Start-time offset requirements for processing the received signals from multiple transmitters in distributed SU MIMO-OFDM under the NLOS case. The figure illustrates an example of the superposition of the transmitted signals from three transmitters (first transmitter with zero time offset, second transmitter with positive time offset, and third transmitter with negative time offset) at the receiver.}
    \label{fig:time_sync_offsets}
    \vspace{-1.75em}
\end{figure}

The effects of residual timing misalignment can be interpreted by considering the LOS and NLOS cases separately. In the LOS case, if all transmitters compensate for their respective propagation delays before CJT, their signals arrive simultaneously at the receiver. In the absence of residual timing offsets, the receiver processing start-time $\Delta$ for applying the Fourier Transform over the useful OFDM symbol interval may be selected anywhere within the CP, i.e., $0 < \Delta < T_{\mathrm{cp}}$.
 In the presence of residual timing offsets, the signals from different transmitters arrive at the receiver at different times, thereby reducing the valid processing interval, where $\Delta_q>0$ indicates that transmitter $q$ starts late and $\Delta_q<0$ indicates that it starts early. This misalignment reduces the receiver processing start-time interval for selecting $\Delta$, which becomes
\begin{equation}
    \max_q(\Delta_q) < \Delta < T_{\mathrm{cp}} + \min_q(\Delta_q).
\end{equation}
In the NLOS case, the delay spread of the channel further shrinks the interval and becomes
\begin{equation}
    \tau^{max} + \max_q(\Delta_q)   < \Delta < T_{\mathrm{cp}} + \min_q(\Delta_q).
    \label{eq:time_interval}
\end{equation}
The $\Delta$ selected within this interval ensures that the timing-induced phase rotations are absorbed into the effective channel and subsequently accounted through precoding. For analytical simplicity, the maximum delay spreads of all
transmitter-receiver links are assumed to be bounded by a common value
$\tau_{\max}$, i.e.,
$
    \tau_{q}^{\max} \leq \tau^{\max},
    \, \forall q \in \{1,\ldots,N_T\}.
$
Fig.~\ref{fig:time_sync_offsets} shows the valid receiver processing start-time
interval, denoted by $T_{\mathrm{margin}}$, in the NLOS case.
Under ideal synchronization, i.e., in the absence of time offsets,
$
T_{\mathrm{margin}} = T_{\mathrm{cp}}-\tau_{\max}.
$
In the presence of residual time offsets, this margin becomes
\begin{equation}
T_{\mathrm{margin}}
=
T_{\mathrm{cp}}+\min_q(\Delta_q)
-\tau_{\max}-\max_q(\Delta_q).
\end{equation}
Therefore, a valid receiver processing interval exists only if
\begin{equation}
\max_q(\Delta_q)-\min_q(\Delta_q) < T_{\mathrm{cp}}-\tau_{\max}.
\end{equation}
Therefore, a time-synchronization technique should minimize the transmitter time-offset spread to limit the required CP overhead.

Using~\eqref{eq:rx_signal_model_per_SC_time_offsets}, the received data on the $k$th subcarrier from all receive antennas at the receiver can be expressed in matrix form as
\begin{equation}
    \mathbf{z}^l_{k}  = \mathbf{G}_k^l
 \FF^l_k \mathbf{d}^l_k  + \mathbf{w}^l_{k} ,
\label{eq:rx_signal_model_per_SC_time_offsets_matrix}
\end{equation}
where ${\mathbf{G}_k^l} = \mathbf{H}^l_{k} [\mathbf{\Phi}^k \otimes \mathbf{I}_{N_t}]$ is the effective channel, $\mathbf{\Phi}^k 
$ = $ \mathrm{diag}\!\Big(\left\{e^{-j2\pi k \phi_q}\right\}_{q=1}^{N_T}\Big)$, and $\phi_q$ = $(T_{\mathrm{cp}}+\Delta_q-\Delta)/T_{\mathrm{eff}}.$
The linear increase of the phase across the subcarriers is different for each transmitter, but remains the same across OFDM symbols for a given transmitter. The Kronecker product in the expression ensures that the phasor term is the same for all transmit antennas within a transmitter. 


\begin{remark}
\textit{If the receiver processing start-time lies within the timing margin and the residual time offsets remain fixed during CE and CJT, the time-offset induced phase rotations are absorbed into the effective channel estimate. Therefore, the precoder can be designed using this effective channel, and the achieved SE matches that under ideal synchronization. }
\end{remark}

Additionally, variations in the transmission start time between CE and CJT, equivalently variations in $\Delta_q$, cause the effective channel during CJT to differ from that used for precoder design. Let $\upsilon_q$ denote the CE-to-CJT start-time jitter of transmitter $q$. We define the sampling-normalized timing jitter as $\bar{\upsilon}_q=F_s\upsilon_q=\upsilon_q/T_s$, where $T_s=1/F_s$ is the sampling period and $\bar{\upsilon}_q$ represents the timing jitter measured in sampling periods. The corresponding transmitter-dependent phase-rotation matrix is
$$
\mathbf{\Xi}^{k}=    \operatorname{diag}\left(    \{e^{-j2\pi k\bar{\upsilon}_q \frac{T_s}{T_{\mathrm{eff}}}} \}_{q=1}^{N_T}\right) = \operatorname{diag}\left(    \{e^{-j2\pi  \frac{k\bar{\upsilon}_q}{K}} \}_{q=1}^{N_T}\right),
$$ 
and the received signal on the $k$th subcarrier of
the $l$th OFDM symbol is
\begin{equation}
\mathbf{z}_{k}^{l} =    \mathbf{G}_{k}^{l} \left[        \mathbf{\Xi}^{k} \otimes        \mathbf{I}_{N_t} \right]    \mathbf{F}_{k}^{l}    \mathbf{d}_{k}^{l}+  \mathbf{w}_{k}^{l}.
    \label{eq:received_signal_time_jitter}
\end{equation}
Thus, the effective channel during CJT is not aligned with the precoder constructed from the channel estimated during CE, resulting in a reduction of coherent combining gain.

\vspace{-1em}
\subsection{Frequency Synchronization Offsets} 
Using~\eqref{eq:rx_signal_model_per_SC_frequency_offsets}, the received data on $k$th subcarrier from all receive antennas at the receiver in matrix form is given by
\begin{align}
    \mathbf{z}^l_{k} = {\mathbf{H}^l_{k} [\mathbf{\Psi}^l \otimes \mathbf{I}_{N_t}]} \FF^l_k \mathbf{d}^l_k + \mathbf{ICI}_k^l + \mathbf{w}^l_{k} ,
\label{eq:rx_signal_model_per_SC_frequency_offsets_matrix}
\end{align}
where $\mathbf{\Psi}^{l} = \mathrm{diag}\!\left( \left\{ \mathrm{sinc}(\psi_q)\, e^{j\pi\psi_q}\, e^{j2\pi\eta_q^{l}} \right\}_{q=1}^{N_T}\right)$, $\psi_q = \delta_q T_{\mathrm{eff}},$ and $\eta_q^{l} = \delta_q (l T_0 + T_{\mathrm{cp}}).$ Here, the $\mathbf{ICI}^l_k$ is a vector of size $N_r \times 1$, expressed as
$$
     \begin{bmatrix}
        \sum_{q,m}^{}   \sum_{i=-\frac{K}{2}, i \neq k}^{\frac{K}{2}-1}s^{m,l}_{q,i} H^{1,m,l}_{q,i} e^{j\pi \chi^{i,k}_q}e^{j2\pi\eta^l_q} \mathrm{sinc}(\chi^{i,k}_q)  \\        \vdots\\    \sum_{q,m}^{}   \sum_{i=-\frac{K}{2}, i \neq k}^{\frac{K}{2}-1}s^{m,l}_{q,i} H^{N_r,m,l}_{q,i} e^{j\pi \chi^{i,k}_q} e^{j2\pi\eta^l_q}\mathrm{sinc}(\chi^{i,k}_q)  \notag     \end{bmatrix}, $$ 
        where $\chi^{i,k}_q = (i-k) + \delta_{q}T_{\mathrm{eff}}$, and it contains the inter-carrier interference (ICI) contributions from each transmit antenna of each transmitter with respect to each receive antenna. This term can be further expressed in matrix form as $$    \mathbf{ICI}^l_k = \sum_{i=-\frac{K}{2}, i \neq k}^{\frac{K}{2}-1}  \mathbf{H}^l_{i} [ \mathbf{X}^{i,k,l} \otimes \mathbf{I}_{N_t}] \FF^l_i \mathbf{d}^l_i,$$ where $\mathbf{X}^{i,k,l}= \mathrm{diag}\!\Big(\left\{\mathrm{sinc}(\chi^{i,k}_q) e^{j \pi \chi^{i,k}_q}e^{j2\pi\eta^l_q}  \right\}_{q=1}^{N_T}\Big).$ 
Compared with the received signal without synchronization offsets in \eqref{eq:rx_signal_subcarrier_matrix}, the additional transmitter-dependent phasor terms $e^{j \pi\psi_q}$ and $e^{j2\pi \eta_q^{l}}$ appear in~\eqref{eq:rx_signal_model_per_SC_frequency_offsets_matrix}, along with ICI. The term $e^{j \pi \psi_q}$ remains constant and is absorbed into the channel response, whereas $e^{j2\pi\eta_q^{l}}$ represents a complex exponential with frequency $\delta_q$ sampled at the OFDM symbol interval $T_0$, introducing a symbol-dependent phase rotation that increases with the OFDM symbol index $l$. Both terms are invariant across subcarriers. In addition, the useful signal power is reduced by a factor of $\norm{\mathrm{sinc}( \psi_q ) }^2$ due to ICI. Since $f_{\mathrm{scs}}=1/T_{\mathrm{eff}}$, $\psi_q$ can also be written as $\delta_q/f_{\mathrm{scs}}$, which denotes the normalized residual CFO. 




\begin{remark}
\textit{The achievable SE decreases relative to the ideal synchronization case due to the useful-signal attenuation caused by the $\mathrm{sinc}(\cdot)$ factor and the ICI. If the residual CFO-induced effective-channel variation is not tracked, the precoder becomes misaligned with the channel, resulting in precoder misalignment and additional SE degradation.}
\end{remark}
%

Combining the effects of residual time and frequency offsets in~\eqref{eq:rx_signal_model_per_SC_time_offsets_matrix} and~\eqref{eq:rx_signal_model_per_SC_frequency_offsets_matrix}, the received data on $k$th subcarrier can be expressed in matrix form as
\begin{align}
    \mathbf{z}^l_{k}  = \mathbf{H}^l_{k} [\mathbf{\Psi}^l \mathbf{\Phi}^k \otimes \mathbf{I}_{N_t}]
 \FF^l_k \mathbf{d}^l_k + \mathbf{ICI}_k^l + \mathbf{w}^l_{k}.
\label{eq:rx_signal_model_matrix_per_SC_time_frequency_offsets}
\end{align}
Here, $\FF^l_k$ should include the phase compensation term that accounts for the residual CFO.

\vspace{-1.25em}
\subsection{Sampling Clock Offsets}
The signal model developed thus far assumes identical sampling clocks at all transmitters and the receiver. Let $\epsilon_q$ denote the residual fractional clock frequency error of transmitter $q$ relative to the receiver. Here, we assume that the sampling clock and carrier frequency of each transmitter are derived from the same reference oscillator~\cite{Merlo2026RealTimeCalibration}. Thus, the same fractional error appears in both the sampling clock and carrier frequency, such that
\begin{equation}
    \epsilon_q
    =
    \frac{F_s^t-F_s}{F_s}
    =
    \frac{f_c^t-f_c}{f_c},
    \label{eq:fractional_clock_error}
\end{equation}
where $F_s^t$ and $F_s$ denote the transmitter and receiver sampling frequencies, respectively, while $f_c^t$ and $f_c$ denote their carrier frequencies. Thus, $\epsilon_q$ also represents the normalized residual sampling clock offset (SCO).  
Errors in the sampling clocks and carrier frequencies can also be modeled independently; however, we leave this as for future extension work. Let $T_s^q=1/F_s^t$ and $T_s=1/F_s$ denote the corresponding sampling intervals. From~\eqref{eq:fractional_clock_error},
\vspace{-.25em}
\begin{equation}
    T_s^q=\frac{T_s}{1+\epsilon_q},
    \qquad
    \epsilon_q=\frac{T_s-T_s^q}{T_s^q}.
    \label{eq:sampling_interval_sco}
\end{equation}
Therefore, the OFDM symbol duration and useful symbol duration at transmitter $q$ are
\vspace{-.25em}
\begin{equation}
    T_0^q=(K+K_{\mathrm{cp}})T_s^q,
    \qquad
    T_{\mathrm{eff}}^q=KT_s^q.
    \label{eq:ofdm_symbol_durations_sco}
\end{equation}
Because the sampling clocks differ across transmitters, their OFDM symbol durations also differ. Consequently, the relative timing offset of transmitter $q$ accumulates over successive OFDM symbols. A positive $\epsilon_q$ corresponds to a faster transmitter clock and shorter OFDM symbols, whereas a negative $\epsilon_q$ corresponds to a slower clock and longer OFDM symbols.

If the accumulated timing misalignment exceeds the timing margin, no feasible value of $\Delta$ exists. Consequently, any chosen value introduces ISI. SCO compensation removes this progressive timing drift. In practice, however, residual SCOs remain and accumulate over time. Therefore, the relative time offsets among the transmitters should be periodically estimated and corrected by adjusting the transmitter times. This periodic correction of time-offsets resets the accumulated timing errors and maintains the received signals at the receiver within the valid timing margin. Additionally, correction must be repeated periodically.



With respect to the receiver sampling interval $T_s$, the sampling interval at transmitter $q$ is
$
    T_s^q
    \approx T_s(1-\epsilon_q),
$
where the approximation holds for $|\epsilon_q|\ll 1$. As an illustrative example, a frequency accuracy error of $1$~ppb corresponds to a fractional SCO of $\epsilon_q=10^{-9}$. For a sampling rate of $15.36$~MHz, the receiver sampling interval is
$
    T_s\approx 65.10~\text{ns}.
$ 
The magnitude of the timing error accumulated per sample is therefore
$
    \approx T_s|\epsilon_q|
    =65.10~\text{ns}\times10^{-9}
    \approx 0.0651~\text{fs}.
    \notag
$
Since this error accumulates approximately linearly with the number of samples, its magnitude after $N$ samples is 
$
    \approx N T_s|\epsilon_q|.
$
Thus, for 1 ppb error, the accumulated timing offset reaches approximately one sampling interval after $10^9$ samples, corresponding to about $65.10$~s.
For transmitter frequency errors bounded within $\epsilon_q\in[-1,+1]$~ppb relative to the receiver, the maximum relative error between any two transmitters is $2$~ppb. This relative error produces a timing separation of approximately two samples after $10^9$ samples. Consequently, the timing margin available for selecting $\Delta$ gradually decreases. Periodically estimating and correcting the relative clock offsets resets the accumulated timing drift and prevents it from exceeding the ISI-free timing margin.

\begin{remark}
\textit{Periodic time-offset correction should be performed before the accumulated relative timing drift exceeds the allowable timing margin. For a correction interval of $L_{\mathrm{r}}$ OFDM symbols, a sufficient condition is
$$
\vspace{-.25em}
    L_{\mathrm{r}} T_0
    (\max_q \epsilon_q - \min_q \epsilon_q)
    \leq T_{\mathrm{margin}}.
    \vspace{-.25em}
$$}
\end{remark}

The received signal model incorporating residual time offsets, SCOs, and CFOs is derived in Appendix~\ref{Derivation5} and can be expressed in matrix form as 
\begin{align}
    \mathbf{z}^l_{k} & = \mathbf{H}^l_{k} [\mathbf{\Psi}^{k,l} \mathbf{\Phi}^k \otimes \mathbf{I}_{N_t}]
 \FF^l_k \mathbf{d}^l_k + \mathbf{ICI}_k^l + \mathbf{w}^l_{k}.
\label{eq:rx_signal_model_matrix_per_SC_time_frequency_sco_offsets}
\end{align}
where $\mathbf{\Psi}^{k,l} = \mathrm{diag}\!\left( \left\{ \mathrm{sinc}(k\epsilon_q + \psi_q)\, e^{j\pi\psi_q}\, e^{j2\pi\eta_{q,k}^{l}} \right\}_{q=1}^{N_T} \right)$, $\mathbf{\Phi}^k  = \mathrm{diag}\!\Big( \left\{e^{-j2\pi k \phi_q}\right\}_{q=1}^{N_T}\Big)$, $\eta_{q,i}^l = (l T_0 + \Delta) (\delta_q+i\epsilon_q/T_{\mathrm{eff}})$, $\psi_q = \delta_qT_{\mathrm{eff}}$, $ \chi_q^{i,k} = i-k+ \delta_q T_{\mathrm{eff}}$, and $\phi_q = -\epsilon_q/2 + (T_{\mathrm{cp}} +\Delta_q - \Delta)/T_{\mathrm{eff}}$. The combined CFO and SCO dependent useful signal attenuation on subcarrier $k$ is captured by SCO term $k\epsilon_q$ and CFO term $\psi_q$. 
Compared with the case involving only normalized CFO, SCO causes additional SE degradation by modifying both the useful signal and ICI terms through $\epsilon_q$. Consequently, subcarriers farther from the center subcarrier are more sensitive to SCO. In addition, the linear phase variation across subcarriers includes an SCO-dependent component through $\phi_q$, while the symbol dependent CFO phase rotation $e^{j2\pi\eta_q^l}$ is changed to symbol- and subcarrier-dependent phase rotation $e^{j2\pi\eta_{q,k}^l}$.

Since the ICI on subcarrier $k$ is a linear combination of the independent zero-mean Gaussian data vectors transmitted on the other subcarriers, it is conditionally complex Gaussian with zero mean and is independent of the desired data vector. Therefore, conditioned on the channel, start-time jitter, time offset, CFO/SCO-offset realizations, its covariance is given by
$\mathbf{R}_{\mathrm{ICI},k}^l=
\mathbb{E}\!\left[\mathbf{ICI}_k^l{\mathbf{ICI}_k^l}^{\hermconj}\right] = $
\vspace{-.25em}
\begin{multline*}
        \sum_{i=-\frac{K}{2}, i \neq k}^{\frac{K}{2}-1} (\mathbf{H}^l_{i} [\mathbf{X}^{i,k,l}\mathbf{\Phi}^i \otimes \mathbf{I}_{N_t}]\left[
        \mathbf{\Xi}^{i}
        \otimes
        \mathbf{I}_{N_t}
    \right] \FF^l_i) \\(\mathbf{H}^l_{i} [\mathbf{X}^{i,k,l} \mathbf{\Phi}^i \otimes \mathbf{I}_{N_t}]    \left[
        \mathbf{\Xi}^{i}
        \otimes
        \mathbf{I}_{N_t}
    \right] \FF^l_i)^{\hermconj}.
\end{multline*}
Thus, the achievable SE is given by 
\begin{equation}
   {R}_{\mathrm{offset}}=  \frac{\kappa}{LK}
    \sum_{l=0}^{L-1}
    \sum_{k=-K/2}^{K/2-1}
    \log_2\det\left(
        \mathbf{I}+\mathbf{\Lambda}^l_{k}
    \right),
    \label{eq:se_offset}
\end{equation}
where
$$ \mathbf{\Lambda}^l_{k} = 
    {\mathbf{R}_{\mathrm{IN},k}^{l}}^{-1/2}
    \widetilde{\mathbf{G}}_{k}^{l}
    \left(
        \widetilde{\mathbf{G}}_{k}^{l}
    \right)^{\hermconj}
    {\mathbf{R}_{\mathrm{IN},k}^{l}}^{-1/2}$$ with $
    \widetilde{\mathbf{G}}_{k}^{l}
    =
    \mathbf{G}_{k}^{l}
    \left[
        \mathbf{\Xi}^{k}
        \otimes
        \mathbf{I}_{N_t}
    \right]
    \mathbf{F}_{k}^{l}
$ and
$ \mathbf{R}_{\mathrm{IN},k}^{l}$ $=$ $\mathbf{R}_{\mathrm{ICI},k}^l
    +\sigma^2\mathbf{I}.
$ 
The effective channel $\mathbf{G}_{k}^l =\mathbf{H}_k^l [\boldsymbol{\Psi}^{k,l}\mathbf{\Phi}^k\otimes\mathbf{I}_{N_t}]$ accounts for the desired signal attenuation and phase rotation caused by the time offsets and CFO/SCO. This assumes that the receiver detects each subcarrier separately, treats the ICI as colored Gaussian noise, and does not perform joint decoding across subcarriers. A joint frequency-domain receiver could instead exploit the complete ICI matrix and potentially achieve a different rate.

\textit{Synchronization-Aware Precoder Design:}
In practical systems, channel estimation (CE) and precoder design are performed prior to CJT by $\mathcal{L}$ OFDM symbols. Let $\mathcal{L}T_0$ denote the feedback delay between CE and CJT. The physical channel is assumed to remain quasi-static over this interval; therefore, the variation in the effective-channel phase is primarily attributed to the residual CFO and SCO. Over a delay of $\mathcal{L}$ OFDM symbols, the transmitter-dependent phase rotation per subcarrier is
$\alpha_{q,k}^{\mathcal{L}}=e^{j2\pi\mathcal{L}T_0 (\delta_q+k\epsilon_q/T_{\mathrm{eff}}) }$. 
With perfect knowledge of this phase rotation, the ideal precoder is given by
\begin{equation}
\widehat{\mathbf{F}}_{k}^{l}
=
\left[
\operatorname{diag}\left(
\left\{
\left(\alpha_{q,k}^{\mathcal{L}}\right)^{*}
\right\}_{q=1}^{N_T}
\right)
\otimes \mathbf{I}_{N_t}
\right]
\mathbf{F}_{k}^{l}.
\end{equation}

In practice, the transmitter-dependent phase rotations must be estimated using a phase-tracking method. Rather than specifying an FDD- or TDD-dependent tracking procedure, we characterize the effect of the phase-estimation accuracy on the SE. The estimated phase rotation is modeled as
\vspace{-.25em}
\begin{equation}
\widehat{\alpha}_{q,k}^{\mathcal{L}}
=
\alpha_{q,k}^{\mathcal{L}}
e^{j\varepsilon_{q,k}^{\mathcal{L}}},
\qquad
\varepsilon_{q,k}^{\mathcal{L}}
\sim\mathcal{N}(0,\sigma_{\varepsilon}^{2}),
\label{eq:phase_estimation_error}
\end{equation}
where $\varepsilon_{q,k}^{\mathcal{L}}$ denotes the phase-estimation error in radians. Accordingly, the precoder based on the estimated tracked-phase is
\begin{equation}
\widetilde{\mathbf{F}}_{k}^{l}
=
\left[
\operatorname{diag}\left(
\left\{
\left(\widehat{\alpha}_{q,k}^{\mathcal{L}}\right)^{*}
\right\}_{q=1}^{N_T}
\right)
\otimes\mathbf{I}_{N_t}
\right]
\mathbf{F}_{k}^{l}.
\label{eq:imperfect_phase_compensating_precoder}
\end{equation}
After compensation, the residual phase rotation associated with transmitter $q$ is
\vspace{-.25em}
\begin{equation*}
\alpha_{q,k}^{\mathcal{L}}\left(\widehat{\alpha}_{q,k}^{\mathcal{L}}\right)^{*} = e^{-j\varepsilon_{q,k}^{\mathcal{L}}}.
\end{equation*}
Thus, $\sigma_{\varepsilon}$ $=$ $0$ corresponds to ideal phase compensation, whereas $\sigma_{\varepsilon}$ $>$ $0$ characterizes imperfect phase tracking, which reduces the coherent-combining gain. Start-time jitter is not modelled here because it is random between CE and CJT and can introduce bias into the phase compensation.


 If the value of $\Delta$ used during CJT reception differs from that used during CE, the received signal can be expressed as
\begin{align*}
    \mathbf{z}^l_{k}  = e^{-j2\pi k\Delta_e/T_{\mathrm{eff}}}{\mathbf{G}_k^l}[\mathbf{\Psi}_k^e \otimes \mathbf{I}_{N_t}]
 \widehat{\mathbf{F}}_{k}^{l}\mathbf{d}^l_k + \mathbf{ICI}_k^l + \mathbf{w}^l_{k},
\vspace{-.25em} 
\end{align*}
where ${\mathbf{G}_k^l} =\mathbf{H}^l_{k} [\mathbf{\Psi}^{k,l} \mathbf{\Phi}^k \otimes \mathbf{I}_{N_t}]$, $\Delta_e$ is the difference between the processing start times used during CE and CJT, and $\mathbf{\Psi}_k^e=\mathrm{diag}\!\Big(\left\{e^{j2\pi  \Delta_e\left[\delta_q+\frac{k\epsilon_q}{T_{\mathrm{eff}}} \right]}\right\}_{q=1}^{N_T}\Big)$. The common subcarrier-dependent phase rotation does not lead to a mismatch between the precoder and the channel. Its estimation and compensation can be performed using precoded pilots, which is a standard approach in communication systems. By contrast, $\mathbf{\Psi}^k_e$ causes a small precoder misalignment; however, its impact is negligible because $\Delta_e$ amounts to only a few sampling intervals, i.e., on the order of tens to hundreds of ns.
\vspace{-1.25em}
\subsection{CSI Feedback Delay}
In practical scenarios, the precoder is designed using the estimated CSI from pilot transmission at time $t-\mathcal{L}T_0$, before CJT at time $t$. The channel may vary between CE and CJT due to the motion of the receiver and scatterers in the environment. This temporal variation is characterized by modeling the channel elements of $\HH_k^l$ in a rich scattering environment as a wide-sense stationary process satisfying $
\mathbb{E}\!\left[H^{n,m,l}_{q,k}\right] = 0,$
\vspace{-.25em}
\begin{align*}
& \mathbb{E}\!\left[\left( H^{n,m,l}_{q,k} \right)^{\hermconj} H^{n',m',l}_{q',k} \right] =
\begin{cases}
1, &
\begin{aligned}
n&=n', \; m=m',\; q=q' \\
\end{aligned} \\
0, & \text{otherwise},
\end{cases} \\
& \rho^{n,m}_{q,k}(\mathcal{L})
=
\mathbb{E}\!\left[
\left( H_{q,k}^{n,m,l} \right)^{\hermconj}
H_{q',k}^{n',m',l-\mathcal{L}}
\right]
=  \\ & \quad \quad\quad\quad\quad\quad
\begin{cases}
J_0\!\left( 2\pi f_d \mathcal{L} T_0 \right), &
\begin{aligned}
n&=n', \; m=m', \; q=q'\\
\end{aligned} \\
0, & \text{otherwise}.
\end{cases}
\end{align*}
where \(J_0\) is the zeroth-order Bessel function of the first kind, and \(f_d\) is the maximum Doppler frequency. For a given feedback delay of $\mathcal{L}$ OFDM symbols, the jointly Gaussian complex channel realizations during CE and CJT
can be related as
\[
\HH_k^l = \rho(\mathcal{L}) \HH_k^{l-\mathcal{L}}
+ \mathbf{E}_k^{\mathcal{L}},
\]
where $\mathbf{E}_k^{\mathcal{L}}$ is independent of
$\HH_k^{l-\mathcal{L}}$ and has independent entries distributed
as $\mathcal{CN}(0,1-|\rho(\mathcal{L})|^2)$.
In the received signal model given in~\eqref{eq:rx_signal_model_matrix_per_SC_time_frequency_offsets},
the precoder $\mathbf{F}_k^l$ is designed using the outdated
effective channel $\mathbf{G}_{k}^{l-\mathcal{L}}$, and thus due to CSI aging the precoder is no longer aligned with the dominant right-singular subspace of the effective channel anymore.
\begin{remark}
   \textit{ CSI aging leads to loss of coherent-combining gain, and this effect worsens with faster CSI aging caused by higher Doppler scenarios and larger feedback delays.}
\end{remark}

\vspace{-1.25em}
\subsection{Phase Noise} The received signal in the presence of phase noise is derived in Appendix~\ref{Derivation6}. From~\eqref{eq:rx_signal_dmimo_ofdm_phase_noise_offsets}, the received data from all receive antennas at the receiver, in the presence of phase noise only, can be expressed in matrix form as
\begin{align}
    \mathbf{z}^l_{k} & = \mathbf{H}^l_{k} [\mathbf{\Omega}^l \otimes \mathbf{I}_{N_t}] \FF^l_k \mathbf{d}^l_k + \mathbf{ICI}_k^l + \mathbf{w}^l_{k} ,
\end{align}
where
$
\mathbf{\Omega}^{l}
=
\mathrm{diag}\!\left(
\left\{
\omega_q^l
\right\}_{q=1}^{N_T}
\right), 
$
Here, the $\mathbf{ICI}^l_k$ is the vector, expressed in matrix form as
\begin{equation*}\mathbf{ICI}^l_k = \sum_{i=-\frac{K}{2}, i \neq k}^{\frac{K}{2}-1} \mathbf{H}^l_{i} [ \mathbf{\Upsilon}^{i,k,l} \otimes \mathbf{I}_{N_t}] \FF^l_i \mathbf{d}^l_i
\end{equation*}
where
$\mathbf{\Upsilon}^{i,k,l}$ = $
\mathrm{diag}\!\left(
\left\{
\upsilon_{q,i,k}^l
\right\}_{q=1}^{N_T}
\right).$
For optimal CJT transmissions, the phase noise should remain steady throughout the CE and CJT intervals so that the pilots accurately estimate the complex common coefficient. In practice, this coefficient varies with time following the PLL’s phase-noise characteristics, and can be different at CJT transmission. Moreover, because the coefficient is the sum of complex exponential phasors, its magnitude is bounded by 1 and may be smaller. As a result, the effective channel gain decreases, which lowers the SNR and consequently degrades the SE.


\begin{remark}
\textit{Phase noise induces a transmitter-dependent complex common coefficient with magnitude less than or equal to one, which evolves randomly between CE and CJT. Thus, the precoder becomes misaligned with the effective channel (also reducing the effective channel gain), with stronger degradation for poorer phase-noise profiles. In addition, phase noise introduces ICI.}
\end{remark}

\begin{figure}[h]
    \centering
    \begin{subfigure}[t]{0.49\columnwidth}
        \centering
\includegraphics[width=.994\columnwidth]{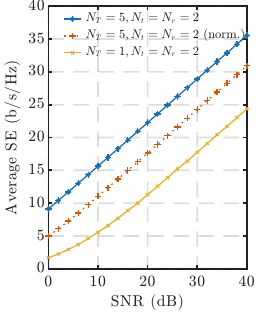}
        \caption{Average SE of CJT for $N_T=5$ transmitters, each with $N_t=2$ transmit antennas, and one receiver with $N_r=2$ receive antennas.}
        \label{fig:cap_ideal}
    \end{subfigure}
    \hfill
    \begin{subfigure}[t]{0.49\columnwidth}
        \centering
    \includegraphics[width=0.994\columnwidth]{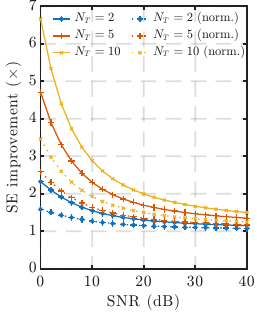}
        \caption{SE improvement of multiple transmitters $N_T = \{2,5,10\}$ over single-transmitter $N_T=1$ SU-MIMO.}
        \label{fig:cap_improvements}
    \end{subfigure}
\caption{Average SE of single-user (SU) DMIMO-OFDM CJT.}   \label{fig:cap_ideal_and_improvement}
\vspace{-1.50em}
\end{figure}

\vspace{-0.5em}
\section{Simulation Results and Analysis} \label{sec:simulation_results_and_analysis}
\vspace{-0.25em}


\noindent We present performance results based on the derived expressions for the single-user DMIMO-OFDM CJT setup with a static multipath Rayleigh fading channel with perfect CSI and phase tracking over 250 Monte Carlo simulation runs. The channel is modeled using independent complex Gaussian taps, generated as $h_{d}\sim\mathcal{CN}(0,1/D_h)$, where $D_h$ is the number of multipath taps, equal to 32. Unless otherwise stated, the simulations consider $N_T=5$ transmitters, each equipped with $N_t=2$ antennas, and a receiver equipped with $N_r=2$ antennas. The OFDM configuration initially uses $K=128$ subcarriers. For a fixed sampling rate, increasing the number of subcarriers under ideal synchronization does not change the SE, because the resulting reduction in subcarrier spacing is accompanied by a proportional increase in OFDM symbol duration (ignored CP overhead). The number of subcarriers is varied in later simulations to evaluate the sensitivity of the system to synchronization offsets. Parameters that differ from these default values are specified for the corresponding results. Additionally, the timing offsets at transmitters are assumed smaller than the timing margin, thus avoiding ISI. 
\begin{figure}[h]
    \centering
    \includegraphics[width=1.0\columnwidth]{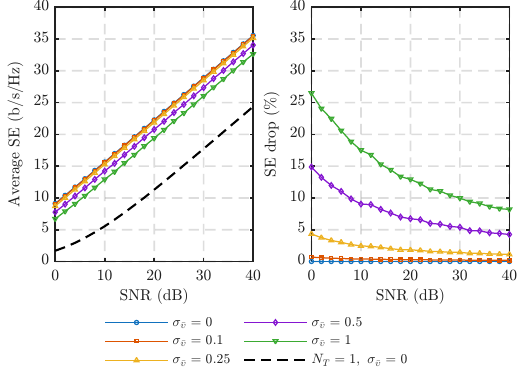}
    \caption{Average SE of SU DMIMO-OFDM CJT under sampling-normalized start-time jitter between CE and CJT (left) and percentage decrease relative to the ideal case (right).}
    \label{fig:time_offset_jitter}
    \vspace{-1.00em}
\end{figure}

Fig.~\ref{fig:cap_ideal} plots the average spectral efficiency (SE) versus $\mathrm{SNR}$ for the five-transmitter case and compares it with the single-transmitter case, both with and without power normalization. Without power normalization, the additional transmit power from each transmitter, along with the increased transmit antennas, leads to higher SE. With power normalization, the SE matches that of a co-located SU-MIMO having the same total antennas as the distributed setup. This behavior changes when path-loss effects are incorporated, where DMIMO outperforms due to the distributed nature of transceivers, resulting in macro-diversity gains. Fig.~\ref{fig:cap_improvements} shows the multiplicative SE gain achieved with additional transmitters $N_T=\{2,5,10\}$ relative to the single-transmitter $N_T=1$ SU-MIMO. The results indicate that the improvement is consistent across SNR in absolute terms, resulting in large multiplicative gains at low SNR. Because SE is strongly dependent on SNR, a transmitter with high SNR will have a larger impact on SE degradation from its synchronization offsets and outdated CSI.



Hereafter, the SE results are presented without power normalization, since the trend of SE degradation caused by offsets remains unchanged. Fig.~\ref{fig:time_offset_jitter} shows the average SE performance for different levels of sampling-normalized start-time CE-to-CJT jitter, modeled as
$\bar{\upsilon}_q\sim\mathcal{N}(0,\sigma_{\bar{\upsilon}}^2)$. As the jitter increases, the SE decreases because the transmitter-dependent phase variations during CJT are not captured by the channel used for precoder design, thereby degrading the coherent-combining gain. \textcolor{black}{The SE degradation remains minimal when the jitter standard deviation is below 10\% of the sampling period ($\approx 6.5$ ns at sampling rate of 15.36 MHz), with the SE loss decreasing from 0.68\% at SNR = 0 dB to 0.17\% at SNR = 40 dB.} 

\begin{figure}[h]
    \centering
    \includegraphics[width=1.0\columnwidth]{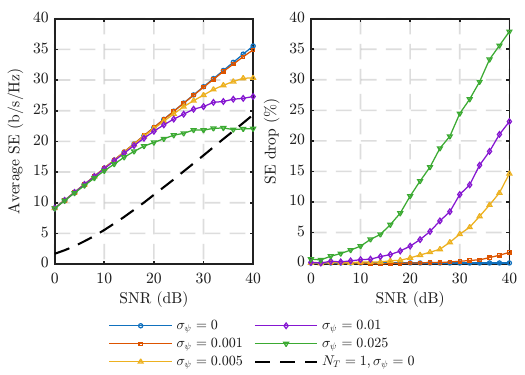}
    \caption{Average SE of SU DMIMO-OFDM CJT under normalized CFO (std.) error $\sigma_{\psi}$ (left) and percentage decrease relative to the ideal case (right).}
    \label{fig:cap_CFO}
    \vspace{-1.25em}
\end{figure}


\begin{figure*}[h]
\centering
\hfill
\begin{subfigure}[t]{0.35\linewidth}
\centering
\includegraphics[width=\linewidth]{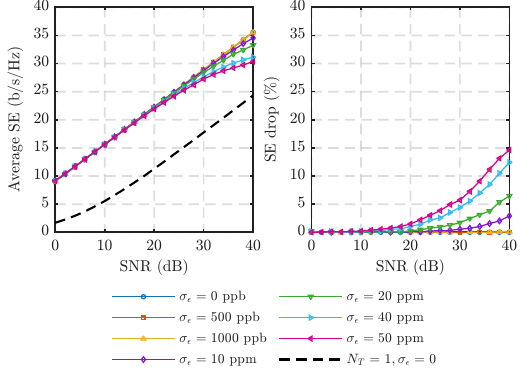}
\caption{$K= 512, f_{\mathrm{scs}}=30$ kHz (SCO only).}
\label{fig:Cap_SCO_only_loss}
\end{subfigure}
\begin{subfigure}[t]{0.35\linewidth}
\centering
\includegraphics[width=\linewidth]{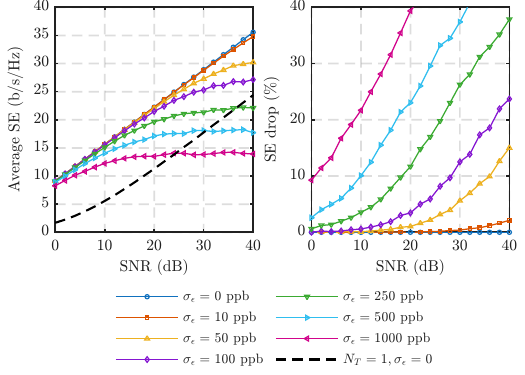}
\caption{$K= 512, f_{\mathrm{scs}}=30$ kHz (SCO \& CFO combined).}
\label{fig:Cap_CSI_SCO_512}
\end{subfigure}
\hfill
\begin{subfigure}[t]{0.28\linewidth}
\centering
\includegraphics[width=1.0\linewidth]{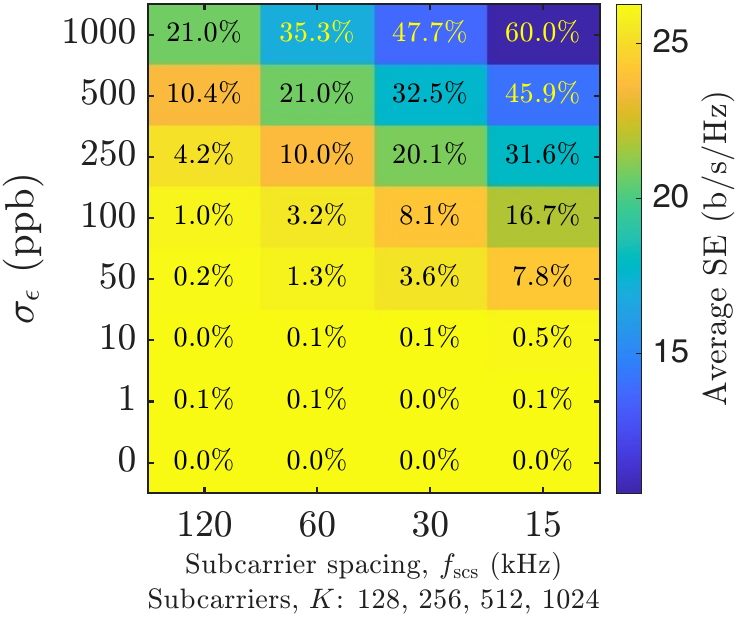} 
\caption{$\mathrm{SNR}=26$ dB; overlaid values show SE loss.}
\label{fig:Cap_CSI_SCO_1024}
\end{subfigure}
\caption{Average SE of SU DMIMO-OFDM CJT under different residual clock frequency errors for $F_s=15.36$ MHz at $f_c=3.5$ GHz.}
\label{fig:capacity_analysis_sco_full_power}
\vspace{-1.50 em}
\end{figure*}

Fig.~\ref{fig:cap_CFO} shows the SE performance under normalized CFO with ideal precoder. The normalized CFO of each transmitter is modeled as $\psi_q\sim\mathcal{N}(0,\sigma_{\psi}^2)$. The results show that increasing the normalized CFO reduces the SE. However, in the low-SNR region, the performance is primarily noise-limited. \textcolor{black}{The results show that a normalized CFO standard deviation of $0.001$ causes negligible ICI-induced degradation, with SE loss increasing from 0.05\% at SNR = 0 dB to 2\% at SNR = 40 dB. For subcarrier spacings of $15$, $30$, $60$, and $120$~kHz, this corresponds to CFO standard deviation of $15$, $30$, $60$, and $120$~Hz, respectively. Additionally, at an SNR of 25 dB, a normalized CFO of 0.005 keeps the loss around 2\%. However, as the SNR approaches 40 dB, the requirement becomes more stringent, with a normalized CFO of 0.001 needed to maintain losses around 2\%.} Nevertheless, CFO-induced phase drift between CE and CJT must be tracked or incorporated into the precoder to preserve coherent gain. Thus, once the normalized CFO standard deviation is below $0.001$, maintaining phase alignment between CE and CJT becomes more important than further reducing the CFO. \textcolor{black}{It also suggests that increasing the subcarrier spacing improves robustness, although it reduces the symbol duration. Thus, the CP duration (assuming a fixed CP-to-symbol duration ratio) and the allowable timing margin for $\Delta$ are reduced. Thus, using a larger CP while ensuring that the associated overhead remains acceptable is needed.}  


\begin{figure}[h]
    \centering
    \includegraphics[width=1.0\columnwidth]{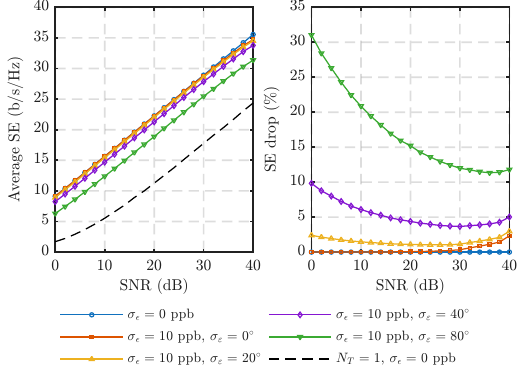}
    \caption{Average SE of SU DMIMO-OFDM CJT under clock-frequency error of 10 ppb and imperfect phase tracking (left) and percentage decrease relative to the ideal case (right).}
    \label{fig:sco_phase_tracking}
    \vspace{-1.25em}
\end{figure}
Fig.~\ref{fig:capacity_analysis_sco_full_power} illustrates the impact of clock frequency error on the SE for a sampling rate of $F_s=15.36$~MHz. The residual CFO and SCO are generated from the clock frequency error, which is modeled as $\epsilon_q\sim\mathcal{N}(0,\sigma_{\epsilon}^2)$, expressed in ppb. Fig.~\ref{fig:Cap_SCO_only_loss} presents the effect of SCO alone, whereas Fig.~\ref{fig:Cap_CSI_SCO_512} presents the combined effect of SCO and CFO at a carrier frequency of $3.5$~GHz. \textcolor{black}{The impact of SCO alone remains negligible up to $1000$~ppb, while the combined degradation remains small up to $10$~ppb, indicating that CFO is the dominant impairment.} Fig.~\ref{fig:Cap_CSI_SCO_1024} presents the SE at $\mathrm{SNR}=26$~dB as a function of the subcarrier spacing (number of subcarriers) and clock error. The overlaid values indicate the SE loss relative to the ideal case with $\sigma_{\epsilon}=0$~ppb. The degradation becomes more noticeable as the subcarrier spacing is reduced because, with a fixed sampling rate, the system becomes more sensitive to synchronization errors. Moreover, for a given clock frequency error, the residual CFO grows with increasing carrier frequency, resulting in more severe ICI and degradation in SE.

While Fig.~\ref{fig:capacity_analysis_sco_full_power} considers an ideal precoder with perfect phase compensation, Fig.~\ref{fig:sco_phase_tracking} evaluates the impact of imperfect phase tracking on the achievable SE for a clock error of $10$ ppb. The residual phase-tracking error is modeled as $\varepsilon_{q,k}^{\mathcal{L}}\sim\mathcal{N}(0,\sigma_{\varepsilon}^2)$, where $\sigma_{\varepsilon}$ denotes the phase-error standard deviation shown in the legend. As $\sigma_{\varepsilon}$ increases, the precoder becomes progressively misaligned with the effective channel, reducing the coherent-combining gain and, consequently, the achievable SE. \textcolor{black}{A phase-error standard deviation of up to $20^\circ$ results in SE degradation around 2.5\%. The loss in coherent-combining gain initially decreases with increasing SNR but grows again at high SNR due to ICI.}

\begin{figure}[h]
    \centering
    \includegraphics[width=0.80\columnwidth]{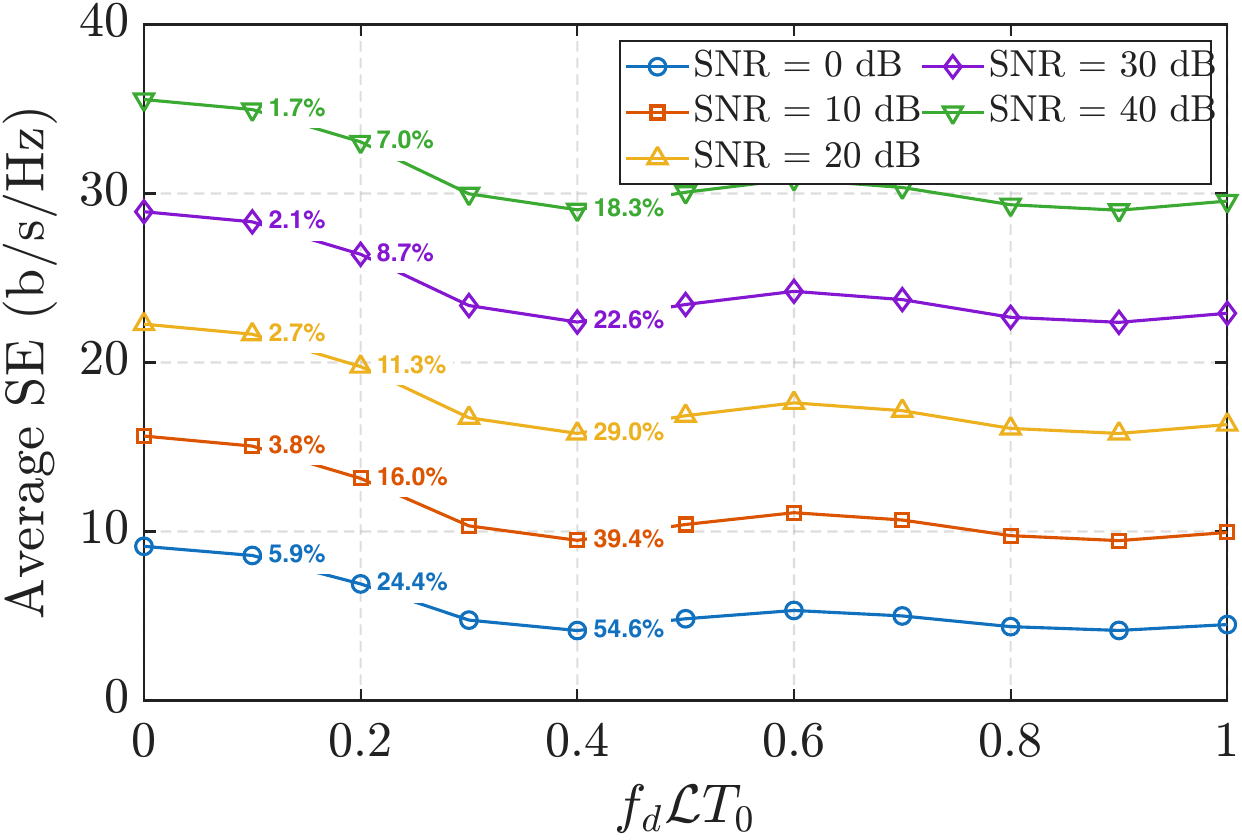}
    \caption{Average SE of SU DMIMO-OFDM CJT at different SNRs as a function of the product of the maximum Doppler frequency and the CSI feedback delay.}
    \label{fig:capacity_analysis_feedback_delay_full_power_csi}
    \vspace{-1.0em}
\end{figure}

Fig.~\ref{fig:capacity_analysis_feedback_delay_full_power_csi} illustrates the SE performance at different SNRs as a function of the product of the maximum Doppler frequency and the CSI feedback delay. A small product corresponds to either a low Doppler frequency or a short feedback delay. As this product increases, the SE degrades because the channel becomes decreasingly correlated between CE and CJT. At larger Doppler-delay products, the oscillatory behavior of the Bessel correlation function produces a slight recovery in SE. \textcolor{black}{The plot also shows the percentage SE degradation at product values of 0.1, 0.2, and 0.4. At an SNR of 30 dB, maintaining the SE degradation below approximately 2\% requires the product to remain below 0.1, while maintaining it below 9\% requires the product to remain below 0.2. For a maximum Doppler frequency of 5 Hz, these conditions correspond to feedback delays below 20 ms and 40 ms, respectively.} The mobility of the receiver and scatterers determines the Doppler characteristics, while the carrier frequency determines the corresponding maximum Doppler frequency. Therefore, under the same mobility conditions, a lower carrier frequency generally results in a more slowly varying channel and, consequently, less CSI aging. Therefore, to mitigate the impact of CSI aging, channel prediction can be employed to design the precoder~\cite{11568206}.

\begin{figure}[h]
    \centering
    \includegraphics[width=1.0\columnwidth]{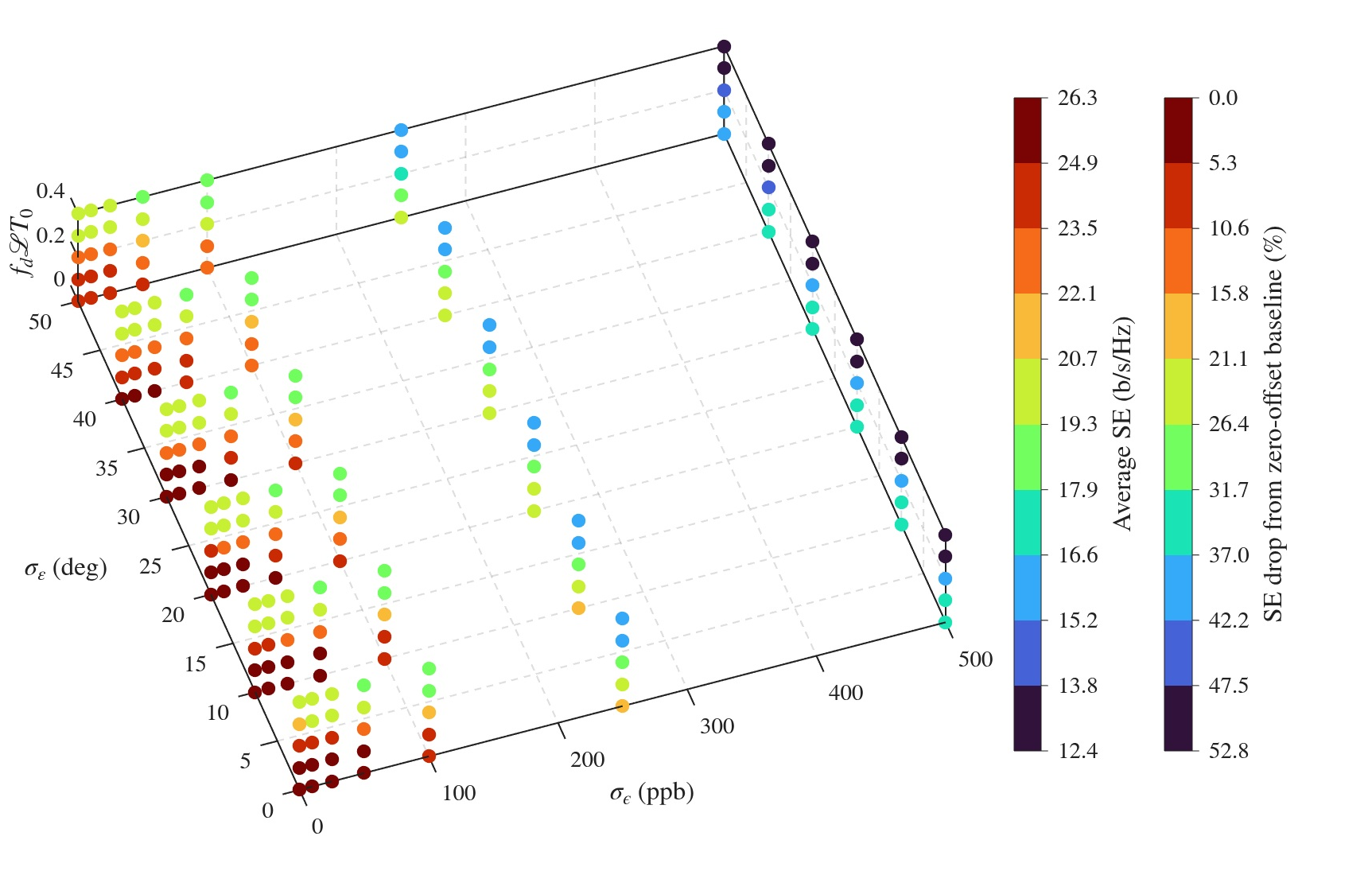}
    \caption{Average SE of SU DMIMO-OFDM CJT under Doppler-delay product, clock errors ($\sigma_\epsilon$), and phase-tracking error ($\sigma_\varepsilon$) at $\mathrm{SNR}=26$ dB.}
    \label{fig:capacity_analysis_feedback_delay_full_power_csi_cfo}
    \vspace{-1.0em}
\end{figure}

Fig.~\ref{fig:capacity_analysis_feedback_delay_full_power_csi_cfo} shows the combined impact of clock error standard deviation (in ppb), which determines the SCO/CFO, phase-tracking error standard deviation, and the Doppler-delay product. The color of each point indicates the average SE, with the two colorbars providing the SE and its percentage reduction relative to the zero-offset baseline at $\mathrm{SNR}=26$~dB. The zero-offset baseline achieves an SE of approximately $26.3$~bit/s/Hz. To maintain an acceptable loss within 5\%, multiple combinations of clock-error standard deviation, phase-tracking error, and Doppler-delay product can be used, resulting in numerous feasible operating points within 50 ppb clock error, a 0.1 Doppler-delay product, and a $40^{\circ}$ phase-tracking error.
\vspace{-0.75em}

\begin{figure}[htpb]
    \centering
    \includegraphics[width=1.0\linewidth]{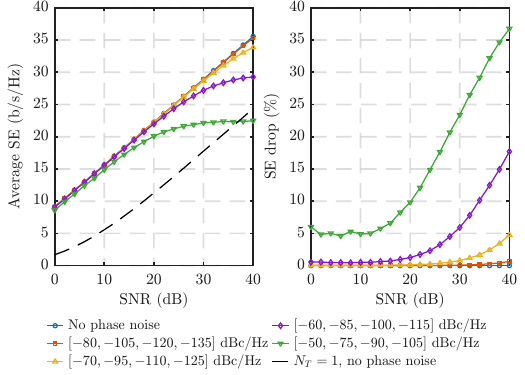 }
    \caption{Average SE of SU DMIMO-OFDM CJT under different phase noise profiles with a feedback delay of $25$~ms, $F_s=15.36$ MHz.}
    \label{fig:phase_noise_sweep}
    \vspace{-1.00em}
\end{figure}

Fig.~\ref{fig:phase_noise_sweep} shows the impact of oscillator phase noise on the SE performance. The phase noise is modeled as filtered Gaussian noise (at the transmitters) and generated using MATLAB \cite{mathworks_phasenoise,381848}, with an assumed baseline profile of $\{-80,\,-105,\,-120,\,-135\}$ dBc/Hz at frequency offsets of $\{1\,\text{kHz},\,10\,\text{kHz},\,100\,\text{kHz},\,1\,\text{MHz}\}$ and a sampling rate of $15.36$~MHz. For comparative evaluation, a uniform offset $\xi \in \{0,10,20,30\}$ dB is applied to the baseline profile. A feedback delay of $25$~ms is considered between CE and CJT, during which the same phase-noise realization evolves continuously and changes the effective channel. This introduces a mismatch between the precoder and the effective channel. As the phase-noise level increases, this mismatch becomes more significant, leading to a loss of coherent-combining gain. 
    
The results show that, as thermal noise becomes less dominant, residual CFO, SCO, and phase-noise impairments, all of which induce ICI, impose a performance ceiling that cannot be overcome by increasing transmit power alone. Consequently, a high received SNR is insufficient to preserve the gains of CJT. This conclusion also applies to practical deployments with path loss and shadowing, which determine the received-SNR operating point but cannot compensate for inadequate synchronization. For CE-to-CJT start-time jitter, phase-tracking errors, and CSI feedback delay, the degradation occurs across the entire SNR range, mainly because the precoder becomes misaligned with the effective channel.

\section{Conclusion and Future Work} \label{sec:conclusion_and_future_work}

\noindent This paper investigated the impact of synchronization offsets and CSI feedback delay on CJT in single-user DMIMO-OFDM systems. Unlike prior work that mainly focused on individual impairments in a single-carrier setting, this work considers wideband multipath propagation and jointly analyzes oscillator-induced impairments and channel time variations. Received signal models were derived to characterize the effects of residual timing offsets, CFO, SCO, Doppler-induced CSI aging and phase noise, thereby providing practical guidance for enabling CJT in distributed deployments. The analysis shows that these impairments degrade CJT performance through precoder misalignment, ICI, and effective channel-gain reduction.  


First, propagation delays must be compensated so that the residual time offset differences among the distributed transmitters remain within the CP. Additionally, these residual timing offsets reduce the available margin within the CP. Separately, CE-to-CJT start-time jitter changes the transmitter-dependent phase between CE and CJT and should therefore be limited to a small sub-sample period. \textcolor{black}{In the considered setup, a Gaussian-distributed start-time jitter with a standard deviation below 10\% of the sampling period results in negligible SE degradation.} 


For frequency synchronization, residual CFO and SCO introduce ICI. The normalized CFO should be kept sufficiently small. \textcolor{black}{A Gaussian-distributed residual CFO with a standard deviation of approximately $0.001$ times the subcarrier spacing causes negligible SE degradation.} The results also indicate that larger subcarrier spacing improves robustness to frequency errors by reducing the normalized CFO. For the same clock error, residual CFO causes greater SE degradation than residual SCO. Although, periodic correction of the transmitter time offsets is required to limit accumulated timing drift due to SCO, and a relationship between correction interval and timing margin is provided. Once the residual timing, CFO, and SCO errors are sufficiently bounded, phase tracking between CE and CJT becomes critical for maintaining precoder alignment. \textcolor{black}{A Gaussian-distributed phase-tracking error with a standard deviation of up to $20^{\circ}$ causes SE degradation around 2.5\%.}



CSI feedback delay causes the precoder to be designed using outdated channel information, preventing it from remaining aligned with the beamforming directions associated with the strongest singular modes. \textcolor{black}{Maintaining the product of the maximum Doppler frequency and feedback delay below 0.1 resulted in an SE loss that decreased from 6\% at an SNR of 0 dB to 1.7\% at 40 dB}. Phase noise also causes precoder misalignment and ICI; however, unlike the deterministic phase rotation caused by residual CFO/SCO, the phase variation is random and depends on the oscillator phase noise profile. Thus, suitable phase noise profiles for the transceiver oscillators are essential to preserve coherent combining gain.

Overall, the analysis shows that practical CJT in DMIMO-OFDM systems requires propagation delay compensation, bounded residual timing, CFO, and SCO errors, reliable phase tracking, and timely CSI feedback. This papers shows gains due to coherent combining can be maintained with reasonable synchronization targets. These impairments should be considered jointly when designing DMIMO systems, since each can reduce the coherent combining gain and limit the achievable SE improvement. Future work will extend this study in two main directions. First, the analysis will be extended to include large-scale path-loss effects and network-level interference using stochastic geometry, since the current study does not explicitly model path loss from spatially distributed transmitters and interferers. Second, the performance and synchronization sensitivity of CJT will be compared with non-coherent joint transmission (N-CJT), which may offer a more practical alternative in high-Doppler environments.

\vspace{-1.00em}

\section*{Acknowledgments}
\noindent Efforts sponsored by the U.S. Government under the Training and Readiness Accelerator II (TReX II), OTA. The U.S. Government is authorized to reproduce and distribute reprints for Governmental purposes, notwithstanding any copyright notation thereon. The views and conclusions contained herein are those of the authors and should not be interpreted as necessarily representing the official policies or endorsements, either expressed or implied, of the U.S. Government.

\vspace{-0.75em}

\appendices
\section{Derivation With Ideal Synchronization}
\label{Derivation1}
Applying the matched filter to~\eqref{eq:received_signal_time} to extract the data on the $k$th subcarrier of the $l$th OFDM symbol, we obtain
\begin{align}
z^{n,l}_{k}
&=
\frac{1}{T_{\mathrm{eff}}}
\int_{lT_0+T_{\mathrm{cp}}}^{lT_0+T_{\mathrm{eff}}+T_{\mathrm{cp}}}
y^n(t)
e^{-j2\pi k(t-lT_0-T_{\mathrm{cp}})/T_{\mathrm{eff}}}dt . \notag
\end{align}
\noindent
Substituting $u=t-lT_0-T_{\mathrm{cp}}$, we can write
\begin{align}
z^{n,l}_{k}
&=
\frac{1}{T_{\mathrm{eff}}}
\sum_{q=1}^{N_T}
\sum_{m=1}^{N_t}
\sum_{i=-K/2}^{K/2-1}
s^{m,l}_{q,i}
\mathcal{I}^{n,m,l}_{q,k,i}
+w^{n,l}_{k}, \notag
\end{align}
\noindent where
{\small
\begin{align}
\mathcal{I}^{n,m,l}_{q,k,i}
&=
\int_{0}^{T_{\mathrm{eff}}}
\int_{0}^{\tau^{\max}_{q}}
h^{n,m,l}_{q}(\tau)
e^{j2\pi i(u-\tau)/T_{\mathrm{eff}}}
e^{-j2\pi ku/T_{\mathrm{eff}}}
d\tau du. \notag
\end{align}
}

\noindent
The channel frequency response on the $i$th subcarrier is
$H^{n,m,l}_{q,i}
=
\int_{0}^{\tau^{\max}_{q}}
h^{n,m,l}_{q}(\tau)
e^{-j2\pi i\tau/T_{\mathrm{eff}}}d\tau .
$
With further simplication, the received data on the $n$th receive antenna, $k$th subcarrier, in the $l$th OFDM symbol of the receiver can be written as 
\vspace{-.25em}
\begin{equation}
    z_{k}^{n,l} = \sum_{q=1}^{N_T} \sum_{m=1}^{N_t} H_{q,k}^{n,m,l} s_{q,k}^{m,l} + w_{k}^{n,l}.
    \label{eq:rx_signal_subcarrier}
\end{equation}
Here, $w_{k}^{n,l}$ denotes the noise at receive antenna $n$ on subcarrier $k$, with variance $\sigma^2$. Let $P_k^l$ denote the transmit power allocated to each transmitter on subcarrier $k$, which is assumed to be identical across all transmitters. Since $\sigma^2$ is the noise variance per receive antenna on each subcarrier, the nominal per-transmitter SNR is defined as
$
\mathrm{SNR}
=\frac{P_k^l}{\sigma^2}.
$


\vspace{-1.25em}
\section{Derivation With Time and Frequency Offsets}
\label{Derivation2}
Applying the matched filter to~\eqref{eq:received_signal_offsets} to extract the data on the $k$th subcarrier of the $l$th OFDM symbol, we obtain
\begin{align}
z^{n,l}_{k}
&=
\sum_{q=1}^{N_T}
\sum_{m=1}^{N_t}
\sum_{i=-K/2}^{K/2-1}
s^{m,l}_{q,i}
H^{n,m,l}_{q,i}
\Phi^{i}_{q}
\Psi^{l}_{q}
\Gamma_{q,k,i}
+w^{l}_{k},
\label{eq:rx_signal_offset_compact}
\end{align}
where $\Psi^{l}_{q} = e^{j2\pi\delta_q(lT_0+\Delta)}$, $ \Phi^{i}_{q} = e^{-j2\pi i(T_{\mathrm{cp}}+\Delta_q-\Delta)/T_{\mathrm{eff}}}$ and 
\begin{align}
\Gamma_{q,k,i}
&=
\frac{1}{T_{\mathrm{eff}}}
\int_{0}^{T_{\mathrm{eff}}}
e^{-j2\pi\left((k-i)u/T_{\mathrm{eff}}
-\delta_q u\right)}du .  \notag
\end{align}
The term $\Delta$ is the processing start time within CP for applying the matched filters. The term $\Gamma_{q,k,i}$ can be written as
\begin{align}
\Gamma_{q,k,i} =
e^{j\pi\left[(i-k)+\delta_qT_{\mathrm{eff}}\right]}
\mathrm{sinc}\left((i-k)+\delta_qT_{\mathrm{eff}}\right). \notag
\end{align}
Separating the desired subcarrier component $(i=k)$ and the ICI terms $(i\neq k)$, we obtain
\begin{align}
z^{n,l}_{k}
&=
\sum_{q=1}^{N_T}
\sum_{m=1}^{N_t}
\underbrace{e^{-j2\pi k(T_{\mathrm{cp}}+\Delta_q-\Delta)/T_{\mathrm{eff}}}}_{I}
\underbrace{e^{j\pi\delta_qT_{\mathrm{eff}}}}_{II}
\underbrace{e^{j2\pi\delta_q(lT_0+\Delta)}}_{III}
\nonumber\\
&\times
H^{n,m,l}_{q,k}
\mathrm{sinc}\left(\delta_qT_{\mathrm{eff}}\right)
s^{m,l}_{q,k}
\nonumber\\
& +
\underbrace{
\sum_{q=1}^{N_T}
\sum_{m=1}^{N_t}
\sum_{\substack{i=-K/2 \\ i\neq k}}^{K/2-1}
\Phi^{i}_{q}
\Psi^{l}_{q}
\Gamma_{q,k,i}
H^{n,m,l}_{q,i}
s^{m,l}_{q,i}
}_{\mathrm{ICI}}
+w^{n,l}_{r,k}.
\label{eq:rx_signal_model_per_SC_time_frequency_offsets}
\end{align}
In the presence of only residual time offsets, the received signal on the $k$th subcarrier is
\begin{align}
z^{n,l}_{k}
&=
\sum_{q=1}^{N_T}
\sum_{m=1}^{N_t}
\underbrace{e^{-j2\pi k(T_{\mathrm{cp}}+\Delta_q-\Delta)/T_{\mathrm{eff}}}}_{I}
H^{n,m,l}_{q,k}
s^{m,l}_{q,k}
+w^{n,l}_{r,k}.
\label{eq:rx_signal_model_per_SC_time_offsets}
\end{align}
The time offsets introduce the transmitter-dependent phasor term $I$, which produces a linear phase variation across subcarriers. Next, considering only residual frequency offsets with $\Delta=T_{\mathrm{cp}}$, the received signal on the $k$th subcarrier is
\small
\begin{align}
&z^{n,l}_{k} = 
\sum_{q=1}^{N_T}
\sum_{m=1}^{N_t}
\underbrace{e^{j\pi\delta_qT_{\mathrm{eff}}}}_{II}
\underbrace{e^{j2\pi\delta_q(lT_0+T_{\mathrm{cp}})}}_{III}
H^{n,m,l}_{q,k}
\mathrm{sinc}\left(\delta_qT_{\mathrm{eff}}\right)
s^{m,l}_{q,k}\notag\\
 & +
\underbrace{
\sum_{q=1}^{N_T}
\sum_{m=1}^{N_t}
\sum_{\substack{i=-K/2 \\ i\neq k}}^{K/2-1}\Gamma_{q,k,i}
e^{j2\pi\delta_q(lT_0+T_{\mathrm{cp}})}
H^{n,m,l}_{q,i}
s^{m,l}_{q,i}
}_{\mathrm{ICI}} +w^{n,l}_{k}.
\label{eq:rx_signal_model_per_SC_frequency_offsets}
\end{align}
\normalsize
\noindent The residual frequency offsets introduce the transmitter-dependent phasor terms $II$ and $III$. The term $II$ produces a constant phase rotation, whereas $III$ produces a symbol-dependent phase rotation across OFDM symbols. 
\vspace{-1em}
\section{Derivation With Sampling Clock Offset}
\label{Derivation5}


Using~\eqref{eq:ofdm_symbol_durations_sco}, the transmitted signal from $m$th antenna of $q$th transmitter can be written as
\begin{align}
x_q^m(t)
=
\sum_p \sum_{i=-K/2}^{K/2-1}
s_{q,i}^{m,p}
e^{j2\pi i(t-T_{\mathrm{cp}}-pT_0^q)/T_{\mathrm{eff}}^q}.
\label{eq:ofdm_sco}
\end{align}
Substituting \eqref{eq:ofdm_sco} into~\eqref{eq:received_signal_offsets}, and defining the channel frequency response as
$H^{n,m,p}_{q,i}$
$=$
$\int_{0}^{\tau^{\max}_{q}}
h_{q}^{n,m,p}(\tau)
e^{-j2\pi i\tau/T_{\mathrm{eff}}^q}d\tau,$
the received signal becomes $y^n(t) =$
\begin{gather}
\sum_{q=1}^{N_T}\sum_{m=1}^{N_t}
\Big(
\sum_p \sum_{i=-K/2}^{K/2-1}
s_{q,i}^{m,p}
H^{n,m,p}_{q,i}
e^{j2\pi i(t-T_{\mathrm{cp}}-pT_0^q)/T_{\mathrm{eff}}^q}\Big) \notag
\\ 
\quad \quad \quad \quad \quad \quad \times e^{j2\pi\delta_q t} +w^{n,p}(t).
\label{eq:rx_signal_cfo_offsets_ofdm_signal}
\end{gather}
\noindent At the receiver, sampling is performed with sampling interval $T_s$, and the sampled received signal can be expressed as $y^n(uT_s)
= $
\begin{align*}
&\sum_{q=1}^{N_T}\sum_{m=1}^{N_t}
\Big(
\sum_p \sum_{i=-K/2}^{K/2-1}
s_{q,i}^{m,p}
H^{n,m,p}_{q,i}
e^{j2\pi i(u(1+\epsilon_q)-K_{\mathrm{cp}}-pK_0)/K} \Big) \notag \\
& \qquad\qquad\qquad\qquad\qquad \times e^{j2\pi \delta_q uT_s}+w^{n,p}(uT_s).
\end{align*}
To extract the data on the $k$th subcarrier on the $l$th OFDM symbol with $\Delta=T_{\mathrm{cp}}$, the matched filter is applied as
\begin{equation*}
z^{n,l}_{k}
=
\frac{1}{K}
\sum_{u=lK_0+ K_{\mathrm{cp}}}^{lK_0+K-1+ K_{\mathrm{cp}}}
y^n(uT_s)
e^{-j2\pi k(u-lK_0 - K_{\mathrm{cp}})/K}.
\end{equation*}
Considering the $p=l$th OFDM symbol and changing the summation index to $u=0,\ldots,K-1$, we obtain $z^{n,l}_{k}=$
\begin{align}
\sum_{q=1}^{N_T}\sum_{m=1}^{N_t}
e^{j2\pi (lT_0 + T_{\mathrm{cp}})\delta_q}
\sum_{i}^{}
s_{q,i}^{m,l}
H^{n,m,l}_{q,i} e^{j2\pi i(l T_0 + T_{\mathrm{cp}})\epsilon_q/T_{\mathrm{eff}}}\notag
\\
 \times
\left(
\frac{1}{K}
\sum_{u=0}^{K-1}
e^{j2\pi u[i(1+\epsilon_q)-k+ \delta_q T_{\mathrm{eff}}]/K}
\right)
+w_{k}^{n,l}. \notag
\end{align}
Defining $
\eta_{q,i}^l = (l T_0 + T_{\mathrm{cp}})(\delta_q+i\epsilon_q/T_{\mathrm{eff}})$, $\psi_q = \delta_qT_{\mathrm{eff}}$, $\psi_q^k=k\epsilon_q +\delta_q T_{\mathrm{eff}}$ and $ \chi_q^{i,k} = i-k+ \delta_q T_{\mathrm{eff}}$, 
and using the finite-sum approximation
$
\frac{1}{K}\sum_{u=0}^{K-1}e^{j2\pi u a/K}
\approx
e^{j\pi a}\mathrm{sinc}(a),
$
the received signal in the presence of residual CFO and SCO becomes
\begin{align*}
& z^{n,l}_{k} =
\sum_{q=1}^{N_T}\sum_{m=1}^{N_t}
e^{j2\pi\eta_{q,k}^l}
e^{j\pi k\epsilon_q} e^{j\pi\psi_q}
H^{n,m,l}_{q,k}
\mathrm{sinc}(\psi_q^k)
s^{m,l}_{q,k} +
\\
& \sum_{q=1}^{N_T}\sum_{m=1}^{N_t}\sum_{\substack{i=-K/2\\ i\neq k}}^{K/2-1}
e^{j2\pi\eta_{q,i}^l}
e^{j\pi i\epsilon_q}
e^{j\pi\chi_q^{i,k}}
H^{n,m,l}_{q,i}\mathrm{sinc}(\chi_q^{i,k})
s^{m,l}_{q,i}\\
& \qquad \qquad \qquad \qquad\qquad \qquad\qquad \qquad\qquad +w_{k}^{n,l}. \notag
\end{align*}
Finally, including start time processing $\Delta$ and the time offset term $\phi_q$ = $-\epsilon_q/2 + (T_{\mathrm{cp}}+\Delta_q-\Delta)/T_{\mathrm{eff}}$, then $\eta_{q,i}^l$ changes to $(l T_0 + \Delta)(\delta_q+i\epsilon_q/T_{\mathrm{eff}})$, the received signal in the presence of time offset, CFO, and SCO is
{\small
\begin{multline}
z^{n,l}_{k}
=
\sum_{q=1}^{N_T}\sum_{m=1}^{N_t}
\underbrace{e^{-j2\pi k\phi_q}}_{I}
\underbrace{e^{j\pi\psi_q}}_{II}
\underbrace{e^{j2\pi\eta_{q,k}^l}}_{III}
H^{n,m,l}_{q,k}
\mathrm{sinc}(\psi_q^k)
s^{m,l}_{q,k}
+\\
\sum_{q=1}^{N_T}\sum_{m=1}^{N_t}
\underbrace{
\sum_{\substack{i=-K/2\\ i\neq k}}^{K/2-1}
e^{-j2\pi i\phi_q}
e^{j\pi\chi_q^{i,k}}
e^{j2\pi\eta_{q,i}^l}
\mathrm{sinc}(\chi_q^{i,k})
H^{n,m,l}_{q,i}
s^{m,l}_{q,i}
}_{\mathrm{ICI}}
\\+ \, w_{k}^{n,l}.
\label{eq:rx_signal_model_per_SC_time_cfo_sco}
\end{multline}
\normalsize}
\vspace{-2.25em}
 \section{Derivation With Phase Noise}
\label{Derivation6}


Adding phase noise $\nu_q^p(t)$ to~\eqref{eq:rx_signal_cfo_offsets_ofdm_signal} and
sampling the received signal with interval $T_s$ gives $y^n(uT_s)
=$
\vspace{-.5em}
\begin{align*}
& \sum_{q=1}^{N_T}\sum_{m=1}^{N_t}
\Bigg(
\sum_p\sum_{i=-K/2}^{K/2-1}
s_{q,i}^{m,p}
H^{n,m,p}_{q,i}
e^{j2\pi i(u(1+\epsilon_q)-K_{\mathrm{cp}}-pK_0)/K} \Bigg) \notag
\\ 
 & \qquad \qquad \qquad \qquad \times
e^{j\left[2\pi\delta_quT_s+\nu_q^p(uT_s)\right]} +w^{n,p}(uT_s) .
\end{align*}
Considering the $p=l$th OFDM symbol, applying the matched filter, and changing the summation index to $u=0,\ldots,K-1$, we obtain
{
\begin{multline}
z^{n,l}_{k}
=
\sum_{q=1}^{N_T}\sum_{m=1}^{N_t}
e^{j2\pi\eta_q^l}
\sum_{i=-K/2}^{K/2-1}
s_{q,i}^{m,l}
H^{n,m,l}_{q,i}
\\
 \times
\left(
\frac{1}{K}
\sum_{u=0}^{K-1}
e^{j\left(
2\pi u[i(1+\epsilon_q)-k+\delta_qT_{\mathrm{eff}}]/K
+\nu_q^l(u)
\right)}
\right)
+w_{k}^{n,l}. \notag
\end{multline}}
To isolate the effect of phase noise, we set the residual CFO/SCO terms to zero. Then, the finite-sum term becomes
\begin{align*}
\frac{1}{K}
\sum_{u=0}^{K-1}
e^{j\left(2\pi u(i-k)/K+\nu_q^l(u)\right)}
&=
\begin{cases}
\omega_q^l, & i=k,\\
\upsilon_{q,i,k}^l, & i\neq k,
\end{cases}
\end{align*}
where
\begin{align}
\omega_q^l
=
\frac{1}{K}
\sum_{u=0}^{K-1}
e^{j\nu_q^l(u)},
\upsilon_{q,i,k}^l
=
\frac{1}{K}
\sum_{u=0}^{K-1}
e^{j\left(2\pi u(i-k)/K+\nu_q^l(u)\right)} . \notag
\end{align}
Here, $\omega_q^l$ represents the complex common coefficient, with $|\omega_q^l|\leq 1$, across all subcarriers of the $l$th OFDM symbol from transmitter $q$, whereas $\upsilon_{q,i,k}^l$ represents the ICI coefficient from subcarrier $i$ to subcarrier $k$. Therefore, the received signal in the presence of phase noise can be expressed as
\begin{multline}
z^{n,l}_{k}
=
\sum_{q=1}^{N_T}\sum_{m=1}^{N_t}
H^{n,m,l}_{q,k}
\omega_q^l
s^{m,l}_{q,k}
\\
+
\sum_{q=1}^{N_T}\sum_{m=1}^{N_t}
\sum_{\substack{i=-K/2\\ i\neq k}}^{K/2-1}
H^{n,m,l}_{q,i}
\upsilon_{q,i,k}^l
s^{m,l}_{q,i}
+w_{k}^{n,l}.
\label{eq:rx_signal_dmimo_ofdm_phase_noise_offsets}
\end{multline}

\vspace{-1.25em}

\bibliographystyle{IEEEtran}
\bibliography{reference}

\end{document}